\documentclass[a4paper,11pt]{article}
\usepackage{jheppub} 
\usepackage{lineno}
\usepackage{appendix}
\usepackage{indentfirst}
\usepackage{xcolor}

\usepackage{cancel}

\arxivnumber{1234.56789} 

\title{\boldmath On $\beta$-function of $\mathcal{N}=2$ supersymmetric integrable sigma models II}

\author[a,b]{Mikhail Alfimov,}
\author[a]{Andrey Kurakin}
\affiliation[a]{HSE University, 6 Usacheva str., Moscow 119048, Russia}
\affiliation[b]{P.N. Lebedev Physical Institute of the Russian Academy of Sciences, 53 Leninskiy pr., Moscow 119991, Russia}

\emailAdd{malfimov@hse.ru}
\emailAdd{kurakin-1993@mail.ru}

\abstract{We continue studying regularization scheme dependence of the $\mathcal{N}=2$ supersymmetric sigma models. In the present work the previous result for the four loop $\beta$-function is extended to the five loop order. Namely, we find the renormalization scheme, in which the fifth loop contribution is completely eliminated, while the fourth loop contribution is represented by the certain invariant, which is coordinate independent for the metrics of some models. These models include complete $T$-duals of the $\eta$-deformed $SU(n)/U(n-1)$ models, as well as $\eta$- and $\lambda$-deformed $SU(2)/U(1)$ models, whose metrics solve the RG flow equation up to the fifth loop order. We also comment on the $\lambda$-deformed $SU(2)/U(1)$ and $SU(3)/U(2)$ case, showing that they satisfy five-loop RG flow equation, and discuss their K\"ahler structure.}

\begin{document}
\maketitle
\flushbottom

\section{Introduction}

In this work, we study two-dimensional sigma models and their renormalization group (RG) flow. The sigma model is a theory in which fields take the values on a (pseudo-)Riemannian manifold $\mathcal{M}$ which is called the target space
\begin{equation}
    X: \Sigma\longrightarrow\mathcal{M}\;.
\end{equation}
where $\Sigma$ stands for $2$-dimensional spacetime. The action of such models can be written as
\begin{equation}
    S[X] = \frac{1}{4\pi}\int \left(G_{ij}(X)\,\partial_aX^i\partial_bX^j h^{ab}\sqrt{|h|} + B_{ij}(X)\,\epsilon^{ab} \partial_a X^i\partial_b X^j\right)d^2x 
\end{equation}
where $ X = (X^1(x), \ldots, X^D(x)) $ represents a set of bosonic fields corresponding to target space coordinates, $G_{ij}(X) $ is the target space metric, $B_{ij}(X)$ is the so-called antisymmetric $B$-field and $h_{ab}$ is the spacetime metric, which will be $\delta_{ab}$ in what follows. The target space metric depends on parameters interpreted as coupling constants.
 
If we restrict ourselves to the case $B_{ij}=0$, i.e. of the zero $B$-field, in its most general form, the renormalization group (RG) flow equation for sigma models can be written as \cite{Ecker:1971xko, Friedan:1980jm}
\begin{equation}\label{eq:rg_flow_eq}
    \dot G_{ij} +\nabla_i V_j + \nabla_j V_i = - \beta_{ij}(G)
\end{equation}
where the dot denotes differentiation with respect to the RG time $t=-\log\Lambda$, and $\Lambda$ is the renormalization scale. The metric is assumed to depend on RG time through the coupling constants. Note that the $\beta$-function appears with a negative sign on the right-hand side. This sign convention can be altered by redefining the sign of the RG time. The vector field $V$ corresponds to coordinate transformations in the target space. Its presence introduces ambiguity in the definition of the $\beta$-function.

The present work is the continuation of previous work made by one of the authors \cite{Alfimov:2021sir,Alfimov:2023evq}, which studies the flow of the renormalization group of the two-dimensional integrable sigma models (see also the study of the relation between one-loop renormalizability and integrability in \cite{Hoare:2020fye}). The main motivation for this study lies in the building of the so-called dual description of these sigma models in terms of the Toda type theories, i.e. the theories determined by the certain set of screening charges.

In previous years, there has been significant progress in this direction. Starting from the renowned article \cite{Fateev:1992tk}, where such a description was obtained for the case of the $\eta$-deformed $O(3)$ sigma model, later in \cite{Fateev:1996ea,Fateev:1996em} it was generalized for the $O(4)$ case. After that, in a series of works \cite{Litvinov:2016mgi,Litvinov:2018bou} it was extended to general $\eta$-deformed $O(N)$ sigma models. Also, it was derived for $\eta$-deformed $\mathbb{CP}^N$ models \cite{Fateev:2019xuq,Litvinov:2019rlv} and several cases of $OSp(N|2m)$ models \cite{Alfimov:2020jpy} with the target space being a deformed supersphere (some related questions for this case were clarified in \cite{Alfimov:2023lkw}). It is important to mention recent progress made for the $O(N)$ models with even $N$ in \cite{Bychkov:2025ftt,Bychkov:2025vex}.

The existence of the dual description mentioned above is based on the key fact that there exist several equivalent systems of screening charges depending on the continuous parameter $b$, which describe the integrals of motion of the considered integrable sigma model. Namely, in the limit $b$ to $0$ these charges define the dual or Toda-type model, whereas in the limit $b$ to $+\infty$ they define the sigma model action. One can see the system of screening charges in the UV limit of the sigma model in question, therefore the study of the RG behaviour of these models becomes important to construct their dual description. However, the fact that the $\beta$-function in \eqref{eq:rg_flow_eq} contains not only Ricci tensor, but also higher curvature corrections, makes this problem very non-trivial.

Despite the one-loop RG flow of the deformed integrable sigma
models mentioned above is relatively well studied, not so much is known about the higher loop corrections in the bosonic case. At two and three loop order this question was studied in \cite{Hoare:2019ark,Hoare:2019mcc}. For the $\eta$- and $\lambda$-deformed $O(3)$ models only it appeared to be possible to find the scheme of renormalization in which the authors of \cite{Alfimov:2021sir} managed to extend the RG equation solution to the $4$-loop order. Nevertheless, it was achieved for the $D=2$ target space only as for higher dimensions the higher-loop contributions to the $\beta$-function contain many different covariant structures consisting of the Riemann, Ricci tensors and covariant derivatives.

In the previous work by one of the authors \cite{Alfimov:2023evq} this problem was addressed for the certain class of deformed integrable sigma models. Namely, for supersymmetric sigma models it appears to be simpler. In the minimal subtraction (MS) scheme the $\beta$-function is zero at the 2nd and the 3rd loop, and the first non-trivial contribution appears at the 4th loop only \cite{Grisaru:1986dk,Grisaru:1986px}. For $\mathcal{N}=1$ theories the result is known for four loops \cite{Grisaru:1986kw} and for $\mathcal{N}=2$ theories even to the fifth loop order \cite{Grisaru:1986wj}. Moreover, the $\mathcal{N}=2$ supersymmetry requires the target space manifold to be K\"ahler \cite{Zumino:1979et}. In the article from the beginning of this paragraph the renormalization scheme was found, in which the four-loop solutions of the RG flow equation for certain integrable deformed sigma models were obtained. To be precise, the complete $T$-dual of the $\eta$-deformed $\mathbb{CP}^N$ metric was shown to satisfy the RG flow equation up to the 4th loop order in this scheme. Together with the knowledge that this metric is K\"ahler \cite{Bykov:2020llx}, it provides an example of the $4$-loop effective action of the $\mathcal{N}=2$ supersymmetric integrable deformed sigma model.

The aim of this paper is to extend the result of \cite{Alfimov:2023evq} to the fifth loop order and to pave the way to find other solutions to the RG flow equation. In order to achieve this goal we further investigate the scheme dependence of the $\beta$-function in the $\mathcal{N}=2$ supersymmetric case, concentrating on the fifth loop order. Besides the $\eta$-deformed metrics, we also investigate the properties of the $\lambda$-deformed ones, paying special attention to the study of their K\"ahler structure. As the tool to do this, we consider the class of solutions, originating from the complete $T$-dual $\eta$-deformed $\mathbb{CP}^N$ metrics, which presumably describe the $\mathcal{N}=2$ SCFTs of Kazama-Suzuki type \cite{Kazama:1988qp}.

Before proceeding with the contents of the article, let us discuss the problem in question in a broader context. Analysis of the higher loop contributions to the $\beta$-function of the $\mathcal{N}=2$ supersymmetric sigma models could be helpful in studying the phenomenon of mirror symmetry. Namely, in the article \cite{Hori:2001ax} there was proven that the duality between the $\mathcal{N}=2$ supersymmetric Liouville theory and the $SL(2,\mathbb{R})/U(1)$ Kazama-Suzuki model, i.e. $\mathcal{N}=2$ supersymmetric fermionic extension of the 2d Black Hole model by obtaining both of them from some gauged linear sigma model (GLSM). The latter side of the duality is derived as the low-energy (IR) limit of the GLSM by utilizing the one-loop RG flow equation. The authors argue that this IR limit is valid to all loops by other means, and therefore taking into account higher loop contributions to the $\beta$-function would serve as an additional justification of the obtained results. The same is true for the supersymmetric extension of the higher rank version for all the models of Kazama-Suzuki type \cite{Kazama:1988qp} of the same duality extensively studied in \cite{Creutzig:2020cmn,Creutzig:2021cyl,Creutzig:2021ykz,Creutzig:2021cem}. Additionally, in \cite{Hori:2001ax} the authors addressed a question how to derive the $\mathcal{N}=2$ supersymmetric sigma models, which are massive, from the action of the specific GLSM. This question was further investigated in \cite{Gupta:2017bcp}. The 2d model derived from GLSM is not exactly the sausage model ($\eta$-deformed $SU(2)/U(1)$ sigma model, but rather a model, which has the shape of the sausage and flows to some $\mathcal{N}=2$ superconformal theory in the IR limit. Therefore, to check this statement in all loop orders one needs to take into account higher loop contributions to the $\beta$-function of the $\mathcal{N}=2$ supersymmetric sigma model. An additional possible generalization is the analysis of the 2d sigma models on non-compact Calabi-Yau backgrounds \cite{Gavrylenko:2023dvp}.

Another interesting direction is to explore the sigma models, which correspond to some CFTs, in the context of compactified dimensions in the superstring theory. In order to maintain the superstring consistency or conformal invariance in other words, we have to require the $\beta$-function of the background metric to vanish. One of the ways to construct physically relevant theory is to compactify six extra dimensions out of ten on some manifold, while keeping the remaining four dimensions to be a flat Minkowski space \cite{Lovelace:1983yv}, then the phenomenological properties will depend on the six-dimensional compact manifold \cite{Candelas:1985en}. At one loop level the vanishing $\beta$-function in both supersymmetric and non-supersymmetric cases requires the manifold to be Ricci-flat. In order to conserve at least one supersymmetry one can organize six extra dimensions to be Calabi-Yau manifold, which is Ricci-flat. However, at higher loops one might expect corrections to these Calabi-Yau backgrounds. As it was discussed above, in the minimal subtraction scheme the first non-zero contributions beyond one-loop order appear at the fourth loop only and, seemingly, these backgrounds have to be corrected. But these corrections are in principle scheme-dependent, and it is important to investigate, whether one can find a renormalization scheme, in which there are no corrections to the above mentioned background.

The paper is organized as follows. In the section 2, we explore the dependence of the $\beta$-function on the choice of regularization scheme. For $\mathcal{N}=2$ supersymmetric sigma models, the target space is a K\"ahler manifold. Consequently, the redefinition of the regularization scheme can be expressed in terms of a local redefinition of the K\"ahler potential. Then we will show that for $\mathcal{N}=2$ supersymmetric sigma models the fifth loop contribution can be eliminated by the choice of the scheme parameters. In the section 3 we will focus on studying the RG flow equation solutions in the chosen renormalization scheme. We study complete $T$-duals to the $\eta$-deformed $\mathbb{CP}^{n-1}$ models and $\lambda$-deformed $SU(n)/U(n-1)$ ones, also exploring their K\"ahler structure for $n=2$ \cite{Sfetsos:2013wia,Hollowood:2014rla} and for $n=3$. In the last section we name all the obtained results and discuss some remaining open questions.

\section{The \texorpdfstring{$\beta$}{beta}-function of the \texorpdfstring{$\mathcal{N}=2$}{N=2} supersymmetric sigma model}

The $\mathcal{N}=2 $ supersymmetry requires the sigma model metric to be K\"ahler \cite{Zumino:1979et} in the absence of the $B$-field. For a K\"ahler manifold, the geometry is described by a scalar function in holomorphic coordinates, the K\"ahler potential $K(\Phi^{\mu}, \bar{\Phi}^{\bar{\nu}})$, which is related to the metric as
\begin{equation}
    G_{\mu\bar\nu}(\Phi,\,\bar\Phi) = \frac{\partial^2 K}{\partial \Phi^{\mu} \partial \bar \Phi^{\bar\nu}}\,,
\end{equation}
where $\Phi,\,\bar\Phi$ 
are superfields, and $X^i$ denotes their bosonic components. In the following, we use Greek indices ($ \mu, \alpha, \ldots, \bar{\nu}, \bar{\beta}, \ldots $) to denote holomorphic and antiholomorphic coordinates respectively, while Latin indices ($ i, j, \ldots $) correspond to real coordinates. The action of $\mathcal{N}=2$ sigma model is given by
\begin{equation}
    S = \frac{1}{4\pi} \int d^2z\, d\theta \,d\bar\theta \;K(\Phi^{\mu},\,\bar\Phi^{\bar\nu})\,,
\end{equation}

For the $\mathcal{N}=2$ supersymmetric sigma model, the renormalization group (RG) flow equation can be expressed in terms of the K\"ahler potential as
\begin{equation}
\dot{K}(t) = -\beta_K(K(t))\;,
\end{equation}
where $t$ is the RG time variable. The $\beta$-function for the $\mathcal{N}=2$ supersymmetric sigma model in the minimal subtraction (MS) scheme has been computed up to the fifth loop in the paper \cite{Grisaru:1986wj}. Non-trivial contributions to the $\beta$-function appear at the first, fourth, and fifth loops. In the MS scheme, the RG flow equation takes the form\footnote{Note that the coefficients for the $n$-loop terms of the $\beta$-function differ by a factor of \(\frac{2}{(4\pi)^n}\) from those quoted in \cite{Grisaru:1986wj}.}
\begin{equation}\label{eq:5flow_equation}
\dot{K}(t) = \frac{1}{2}\log\det G - \Delta K - \beta_K^{(5)} + \mathcal{O}(\hbar^5)\;,
\end{equation}
where we assume that the metric $G_{\mu\bar{\nu}}$ is of the order $\hbar^{-1}$ and $\Delta K$ is the fourth-loop contribution to the $\beta$-function, defined as\footnote{The $\Delta K$ introduced here differs from \cite{Grisaru:1986px} by a factor of $-\frac{\zeta(3)}{12}$.}
\begin{align}\label{DeltaK}
\Delta K &= -\frac{\zeta(3)}{12}{{{{R^{\mu}}_{\nu}}^{\tau}}_{\lambda}}\left({{{{R^{\rho}}_{\mu}}^{\alpha}}_{\tau}}{{{{R^{\nu}}_{\rho}}^{\lambda}}_{\alpha}} + {{{{R^{\lambda}}_{\tau}}^{\rho}}_{\alpha}}{{{{R^{\alpha}}_{\rho}}^{\nu}}_{\mu}} \right) = \\
&= \frac{\zeta(3)}{24}R_{ijkl}{{{R^i}_{mn}}^l}\left(R^{jnmk} + R^{jkmn}\right), \notag
\end{align}
and the fifth-loop contribution $\beta_K^{(5)}$ is given by\footnote{The Ricci tensor we will define as the contraction of the first and third indices of the Riemann tensor
\begin{equation}
    R_{\alpha\bar\beta} = {R^{\gamma}}_{\alpha\gamma\bar\beta}\;.
\end{equation}}
\begin{align}\label{eq:5loopbeta}
\beta_K^{(5)} &= -\frac{3\,\zeta(4)}{10\,\zeta(3)} \left(-\nabla^2\Delta K + \right. \\
&\left.+\frac{\zeta(3)}{4}\left(\nabla_{\rho}\nabla^{\sigma}{R_{\nu}}^{\mu} - {R_{\rho}}^{\pi}{{{R_{\pi}}^{\sigma}}_{\nu}}^{\mu} \right)\left({{{R_{\sigma}}^{\alpha}}_{\mu}}^{\tau} {{{R_{\alpha}}^{\rho}}_{\tau}}^{\nu} + {{{R_{\mu}}^{\nu}}_{\alpha}}^{\tau}{{{R_{\tau}}^{\alpha}}_{\sigma}}^{\rho}\right) \right)\;. \notag
\end{align}
The key observation about the second line of \eqref{eq:5loopbeta} is that this formula can be understood as follows
\begin{equation}
    \frac{\zeta(3)}{4}\left(\nabla_{\rho}\nabla^{\sigma}{R_{\nu}}^{\mu} - {R_{\rho}}^{\pi}{{{R_{\pi}}^{\sigma}}_{\nu}}^{\mu} \right)\left({{{R_{\sigma}}^{\alpha}}_{\mu}}^{\tau} {{{R_{\alpha}}^{\rho}}_{\tau}}^{\nu} + {{{R_{\mu}}^{\nu}}_{\alpha}}^{\tau}{{{R_{\tau}}^{\alpha}}_{\sigma}}^{\rho}\right)=\left.\frac{d}{dt}\Delta K\right|_{\dot{G}_{\mu\nu}=-R_{\mu\nu}}\,,
\end{equation}
namely, we can understand it as a derivative of \eqref{DeltaK} with respect to the RG time assuming that the metric satisfies one-loop RG flow equation. Let us further denote this contribution as just $\mathcal{D}\Delta K$.Thus, the expression for $\beta_K^{(5)}$ can be rewritten in a more compact form using $\Delta K$
\begin{equation}\label{eq:beta5}
\beta_K^{(5)} = -\frac{3\,\zeta(4)}{10\,\zeta(3)} \left( -\nabla^2\Delta K + \mathcal{D}\Delta K \right)\,.
\end{equation}

Now let us turn to the question how to make the $\beta$-function simpler by changing the renormalization scheme, or, in other words, find some \textit{normal} form. Such a change can be implemented through a covariant redefinition of the metric \cite{Friedan:1980jm}. However, arbitrary redefinition can make the metric non-K\"ahler. To avoid this, it is convenient to work instead with the redefinition of the K\"ahler potential, in order for the resulting redefined metric to be automatically K\"ahler. Such redefinitions were described in \cite{Nemeschansky:1986yx} in the context of supersymmetric sigma models on Calabi-Yau manifolds, however, they were in principle non-local. We consider the local ones and, in general, as was already mentioned in \cite{Alfimov:2023evq}, a covariant change of scheme may be written as\footnote{As it was explained in \cite{Alfimov:2023evq}, one can rewrite the corresponding covariant metric redefinition as a covariant function of metric and its derivatives in arbitrary non-holomorphic coordinates.}
\begin{align}\label{Kaehler_potential_redefinition}
K \rightarrow \tilde{K}(K) &= K + c_1 \log\det G + c_2 R + \underbrace{c_3 R^2 + c_4 (R_{ijkl})^2 + c_5 (R_{ij})^2 + c_6 \nabla^2 R}_{\text{scalars } \sim R^2} \\
&+ \underbrace{c_7 R^3 + \ldots + c_{23} \nabla^2\nabla^2 R}_{\text{scalars } \sim R^3} + \ldots\;, \notag
\end{align}
where the number of available scalar invariants grows rapidly with the curvature order, although many of them are not independent. In particular, K\"ahler geometry provides additional identities that further reduce the set of independent scalars (see details in appendix \ref{app_Kaehler_identities}). As the result, the RG flow equation in terms of the redefined potential \(\tilde{K}\) becomes
\begin{equation}\label{eq:RG_flow_equation2}
\dot{\tilde{K}}(t) = -\tilde{\beta}_K(\tilde{K}(t))\;,
\end{equation}
where $\tilde{\beta}_K(\tilde{K})$ may take a simpler form than $\beta_K(K)$ in the MS scheme.

It is the right moment to point out an observation, which would be useful in what follows. Based on the form of $\beta_K^{(5)}$ in \eqref{eq:beta5}, it is possible to see that modifying the regularization scheme can eliminate the fifth-loop contribution to the RG flow equation. To achieve this, we modify the regularization scheme by these means
\begin{equation}\label{eq:changing1}
    K \rightarrow  K'(K) = K + c\, \Delta K = K - c\, \, \frac{\zeta(3)}{12}\,{{{{R^{\mu}}_{\nu}}^{\tau}}_{\lambda}}\left({{{{R^{\rho}}_{\mu}}^{\alpha}}_{\tau}}{{{{R^{\nu}}_{\rho}}^{\lambda}}_{\alpha}} + {{{{R^{\lambda}}_{\tau}}^{\rho}}_{\alpha}}{{{{R^{\alpha}}_{\rho}}^{\nu}}_{\mu}} \right)\;,
\end{equation}
where
\begin{equation}
    c = \frac{3\,\zeta(4)}{10\,\zeta(3)}\;.
\end{equation}
Therefore, we have found the modification of the MS regularization scheme, in which the fifth-loop contribution to the $\beta$-function vanishes, and the RG flow equation takes the form \eqref{eq:5flow_equation2}
\begin{equation}\label{eq:5flow_equation2}
    \dot K' = \frac{1}{2}\log\det G' - \Delta K(G') + \mathcal{O}(\hbar^5)\;.
\end{equation}

The main objective of this section is to identify the regularization scheme in which the higher–loop contributions to the $\beta$-function become simpler and we could hope, that we can find the solutions of the RG flow equation up to the fifth loop order. It was demonstrated in \cite{Alfimov:2023evq} that an appropriate redefinition of the scheme allows one to express the fourth loop contribution to the $\beta$-function in terms of certain invariant given by the following formula in complex coordinates
\begin{align}\label{DeltatildeK}
\Delta \tilde{K} &= \frac{\zeta(3)}{12} \left(-{{{{R^{\mu}}_{\nu}}^{\tau}}_{\lambda}}\left({{{{R^{\rho}}_{\mu}}^{\alpha}}_{\tau}}{{{{R^{\nu}}_{\rho}}^{\lambda}}_{\alpha}} + {{{{R^{\lambda}}_{\tau}}^{\rho}}_{\alpha}}{{{{R^{\alpha}}_{\rho}}^{\nu}}_{\mu}} \right) - \right. \\
&\left.-3 R^{\alpha\bar{\beta}} \nabla_{\alpha} \nabla_{\bar{\beta}} R - 5 R^{\alpha\bar{\beta}} R_{\alpha\bar{\gamma}} {R^{\bar{\gamma}}}_{\bar{\beta}} + \frac{3}{2} \nabla^2 R_{\alpha\bar{\beta}}^2 \right)\;. \notag
\end{align}
Moreover, in that work it was shown that for certain classes of metrics, such as complete $T$-dual to the $\eta$-deformed $\mathbb{CP}^{n-1}$ model, including the sausage model with its $T$-dual and $\lambda$-deformed $\mathbb{CP}^1$ model, this invariant \eqref{DeltatildeK} does not depend on the coordinates. Before proceeding with the arbitrary dimension of the target space, let us consider a simpler case of two dimensions.

\subsection{\texorpdfstring{$D=2$}{D=2} target space}

As a preliminary step, let us consider a simpler setting -- two–dimensional target spaces. In two dimensions the expressions simplify substantially: every scalar term can be expressed solely through the curvature and its covariant derivatives. For example, at the order $\sim R^3$ there exist only four independent scalars
\begin{equation}
    R^3\,,\; R\nabla^2R\,,\; \nabla_i R \nabla^i R\,,\; \nabla^2\nabla^2R\,.
\end{equation}
One can also exploit the fact that the K\"ahler potential is defined only up to addition of an arbitrary holomorphic and antiholomorphic functions. In particular, adding to the K\"ahler potential any invariant independent of the target space coordinates leaves the metric unchanged. The invariant
\begin{equation}\label{eq:2dimInv}
    \Delta\tilde K=\frac{\zeta(3)}{12}\left(-\frac{3}{8}R^3-\frac{3}{8}R\nabla^2 R+\frac{3}{8}\nabla_i R\nabla^i R\right)\,,
\end{equation}
does not depend on the coordinates for the $\eta$-deformed $\mathbb{CP}^1$, its $T$-dual and $\lambda$-deformed $SU(2)/U(1)$ model. The presence of the invariant $\Delta\tilde K$ provides additional freedom in specifying the regularization scheme, since for these backgrounds it satisfies the conditions
\begin{equation}
\nabla^2 \Delta\tilde{K}=0\;, \quad \nabla^2 \nabla^2 \Delta\tilde{K}=0, \quad \textrm{etc.} 
\end{equation}
and  the following terms are also coordinate independent and do not contribute to the metric $\beta$-function
\begin{equation}
    \mathcal{D}\Delta \tilde K \,,\quad \mathcal{D}^2\Delta \tilde K \, , \quad \textrm{etc.} 
\end{equation}

In particular, the invariant of order $R^4$ ($\sim\hbar^4$) discussed in the Conclusions section of \cite{Alfimov:2023evq} turns out to be proportional to $\mathcal{D}\Delta\tilde{K}$ 
\begin{align}
    \mathcal{D}\Delta\tilde{K} &\sim R^4 +3R^2\nabla^2R - \frac{1}{6} R\nabla^2 R^2 - \frac{1}{9}\nabla^2 R^3+ \\
    &+\frac{2}{3}\left((\nabla^2 R)^2+R\nabla^2\nabla^2R \right) -\frac{1}{3}\nabla^2(R\nabla^2R)\;. \notag
\end{align}

Now, let us utilize the following idea. Our task at the moment is to adjust the regularization scheme in a way that for the $D=2$ backgrounds mentioned in this subsection the $\beta$-function at the 4th and 5th loops is zero. From \cite{Alfimov:2023evq} we know how to achieve this at the 4th loop. However, the scheme redefinition described in that article would result in some non-zero contribution at the 5th loop order $\beta$-function. Therefore, we have to adjust the scheme redefinition parameters at the second row of \eqref{Kaehler_potential_redefinition} to compensate this contribution. 

This can be achieved in two steps. First, as it was shown previously, we can redefine the MS scheme in order to eliminate the fifth loop contribution as in \eqref{eq:5flow_equation2}, getting
\begin{equation}
    \beta'_{K}=-\frac{1}{2}\log\det G'+\Delta K(G')+\mathcal{O}(\hbar^5)\;.
\end{equation}
Second, we want to find the scheme redefinition
\begin{align}\label{intermediate_scheme_redefinition1}
K' \rightarrow \tilde{K}(K')&=K+c_1 \log\det G'+c_2 R+c_3 R_{ij}R^{ij}+c_4 \nabla^2 R+ \\
&+b_5 R^3+b_6 R\nabla^2 R+b_7 \nabla_i R\nabla^i R+b_8 \nabla^2 \nabla^2 R\,,\notag
\end{align}
where the scalar curvatures and covariant derivatives are calculated for the metric $G'$ and which turns the $\beta$-function into the form
\begin{equation}\label{intermediate_scheme_beta1}
    \tilde{\beta}_K=-\frac{1}{2}\log\det \tilde{G}+\Delta\tilde{K}(\tilde{G})+q_1 \nabla^2 \Delta\tilde{K}+q_2 \mathcal{D}\Delta\tilde{K}+\mathcal{O}(\hbar^5)
\end{equation}
for some $q_1$ and $q_2$. Indeed, this can be done by the following choice of the parameters
\begin{align}\label{scheme_parameters_choice1}
c_2 &= -c_1^2\;, \quad c_3 = -\frac{2c_1^3}{3}+\frac{5\zeta(3)}{48}\;, \quad c_4 = -\frac{c_1^3}{3}+\frac{\zeta(3)}{48}\;, \\
b_5 &= -\frac{c_1^4}{6}\;, \quad b_6 = -\frac{5c_1^4}{12}+\frac{5\zeta(3)}{48}c_1\;, \quad b_7 = -\frac{c_1^4}{6}+\frac{5\zeta(3)}{48}c_1, \quad b_8 = -\frac{c_1^4}{12}+\frac{\zeta(3)}{48}c_1\,. \notag
\end{align}
Given \eqref{scheme_parameters_choice1} and $q_1=c_1$ and $q_2=2c_1$, we obtain \eqref{intermediate_scheme_beta1}. However, this additional contribution can be nicely absorbed by a slight modification of the initial scheme redefinition
\begin{align}\label{intermediate_scheme_redefinition2}
K' \rightarrow \tilde{K}(K')&=K+c_1 \log\det G'+c_2 R+c_3 R_{ij}R^{ij}+c_4 \nabla^2 R+ \\
&+b_5 R^3+b_6 R\nabla^2 R+b_7 \nabla_i R\nabla^i R+b_8 \nabla^2 \nabla^2 R+b_9 \Delta\tilde{K}\,,\notag
\end{align}
which, after adding $b_9=c_1$ allows to eliminate the terms in \eqref{intermediate_scheme_beta1}, proportional to $q_1$ and $q_2$, yielding
\begin{equation}
    \label{intermediate_scheme_beta2}
    \tilde{\beta}_K=-\frac{1}{2}\log\det \tilde{G}+\Delta\tilde{K}(\tilde{G})+\mathcal{O}(\hbar^5)\;.
\end{equation}
The last observation is that we can combine the change of the scheme \eqref{intermediate_scheme_redefinition2} with the change of the scheme \eqref{eq:changing1}, resulting the following change of regularization scheme
\begin{align}\label{overall_scheme_redefinition}
K \rightarrow \tilde{K}(K) &= K + c_1 \log\det G + c_2 R + c_3 R_{ij}R^{ij} + c_4\nabla^2 R + \\
&+ b_5 R^3 + b_6 R \nabla^2 R + b_7 \nabla_i R \nabla^i R + b_8 \nabla^2 \nabla^2 R + b_9 \Delta \tilde{K}\,,\notag
\end{align}
with the coefficients
\begin{align}\label{scheme_parameters_choice2}
c_2 &=-c_1^2\;, \quad c_3=-\frac{2c_1^3}{3}+\frac{5\zeta(3)}{96}\;, \quad c_4=-\frac{c_1^3}{3}+\frac{\zeta(3)}{48}\;, \quad b_5=-\frac{c_1^4}{6}+\frac{\zeta(4)}{160}\;, \\
b_6 &= -\frac{5c_1^4}{12}+\frac{5\zeta(3)}{48}c_1\;, \quad b_7=-\frac{c_1^4}{6}+\frac{5\zeta(3)}{48}c_1\;, \quad b_8=-\frac{c_1^4}{12}+\frac{\zeta(3)}{48}c_1\;, \quad b_9=-2c_1\;. \notag
\end{align}
Then, by direct calculation, we have shown that, in the $D=2$ target space, starting from the MS scheme, in which the RG flow equation has the form \eqref{eq:5flow_equation}, by applying scheme redefinition, given by \eqref{overall_scheme_redefinition} and \eqref{scheme_parameters_choice2}, the potential $\tilde K$ satisfies the equation
\begin{equation}\label{eq:newSchemeEq}
\dot{\tilde{K}} = \frac{1}{2} \log\det \tilde{G} - \Delta \tilde{K} + \mathcal{O}(\hbar^5)\;.
\end{equation}
Note that there remains a residual freedom in the choice of regularization scheme, since the coefficient $c_1$ is arbitrary. The objective of the next subsection is to generalize this result to the case of the target space with arbitrary dimension.

\subsection{Arbitrary \texorpdfstring{$D$}{D} target space}

We now turn to the case of a target space of arbitrary dimension. The situation becomes different as in the scheme redefinition \eqref{Kaehler_potential_redefinition} the number of covariant structures is not reduced by the fact that the target space is two-dimensional. Nevertheless, by an appropriate choice of regularization scheme, the non-trivial contributions of the fourth and fifth loop orders to the $\beta$-function can likewise be made trivial at least for the certain types of sigma models analyzed in the present article, i.e. complete $T$-dual $\eta$-deformed $\mathbb{CP}^{n-1}$ models and $\lambda$-deformed $\mathbb{CP}^1$ together with its $\eta$-deformed version and its $T$-dual, whereas for other $\mathcal{N}=2$ supersymmetric sigma models it remains the subject of further investigation. The number of covariant structures in the change of scheme \eqref{Kaehler_potential_redefinition} is still reduced by the fact that we are dealing with the K\"ahler manifold and thus Riemann, Ricci tensors and covariant derivatives are subject to certain relations (see appendix \ref{app_Kaehler_identities} for details)
\begin{align}\label{eq:regul_scheme5}
K \rightarrow \tilde{K}(K) = K + c_1 \log\det G + c_2 R + 2c_3 (R_{\alpha\bar{\beta}})^2 + 2c_4 \nabla_{\alpha}\nabla^{\alpha} R +\\
+c_5 R_{\alpha\bar{\beta}} R^{\gamma\bar{\beta}} {R^{\alpha}}_{\gamma} + 
c_6 R^{\alpha\bar{\beta}} \nabla_{\gamma}\nabla^{\gamma} R_{\alpha\bar{\beta}} + c_7 R^{\alpha\bar{\beta}} \nabla_{\alpha}\nabla_{\bar{\beta}} R + \notag
\\
+c_8 \nabla_{\gamma} R_{\alpha\bar{\beta}} \nabla^{\gamma} R^{\alpha\bar{\beta}}+c_9 \nabla_{\alpha}\nabla^{\alpha}\nabla_{\beta}\nabla^{\beta} R + c_{10} \Delta K + c_{11} \Delta\tilde{K}\;,\notag
\end{align} 
where the last two terms are added for convenience and $\Delta \tilde{K}$ from \eqref{DeltatildeK} can be concisely written as
\begin{equation}\label{eq:deltaTildeK}
\Delta \tilde{K} = \Delta K + \frac{\zeta(3)}{12} \left( -3 R^{\alpha\bar{\beta}} \nabla_{\alpha} \nabla_{\bar{\beta}} R - 5 R^{\alpha\bar{\beta}} R_{\alpha\bar{\gamma}} {R^{\bar{\gamma}}}_{\bar{\beta}} + \frac{3}{2} \nabla_{\gamma}\nabla^{\gamma} R_{\alpha\bar{\beta}}^2 \right),
\end{equation}
and the numerical coefficients in the first line of \eqref{eq:regul_scheme5} are included in order to disappear after returning from the holomorphic to the real coordinates.
The coordinate independence of \eqref{eq:deltaTildeK} has been verified for the complete $T$-dual of the $\eta$-deformed $\mathbb{CP}^{n-1}$ model with $n=2,4,8$, as well as for the $\lambda$-deformed $SU(2)/U(1)$ and $SU(3)/U(2)$ models in \cite{Alfimov:2023evq}. In the next section we will explicitly find the value of this invariant for the $\lambda$-deformed $SU(3)/U(2)$ model.

In general, computation of the $\beta_K$-function modification generated by \eqref{eq:regul_scheme5} is a hard problem. The details of the calculation are written in the appendix \ref{app_scheme_change}, so we are not going to write it down here as they are too cumbersome. We found, that if we set the coefficients in the modified scheme as
\begin{align}\label{scheme_c_parameters}
c_2 &= -c_1^2\;, \quad c_3 = -\frac{2 c_1^3}{3} + \frac{5 \zeta(3)}{48}\;, \quad c_4 = -\frac{c_1^3}{3} + \frac{\zeta(3)}{48}\;, \quad c_5 = -\frac{4c_1^4}{3}\;, \\
c_6 &= -\frac{4c_1^4}{3}+\frac{5\zeta(3)}{6}c_1\;, \quad c_7 = -c_1^4\;, \quad c_8 = -\frac{4c_1^4}{3}+\frac{5\zeta(3)}{6}c_1\;, \notag \\
c_9 &= -\frac{c_1^4}{3}+\frac{\zeta(3)}{12}c_1\;, \quad c_{10} = \frac{3 \zeta(4)}{10 \zeta(3)}, \quad c_{11} =-2c_1,\notag
\end{align}
then for the potential $\tilde K$ the RG flow equation takes the  form \eqref{eq:newSchemeEq}. Let us note that if we confine ourselves to the case of $D=2$ target space, then the $b$-coefficients from \eqref{scheme_parameters_choice2} are related to the parameters \eqref{scheme_c_parameters} in the following way
\begin{align}
    b_5&=\frac{c_5}{8}+\frac{\zeta(3)}{48}c_{10}\;, \quad b_6=\frac{c_6}{8}+\frac{c_7}{4}\;, \quad b_7=\frac{c_8}{8}\;, \quad b_8=\frac{c_9}{4}\;.
\end{align}

In this way we obtain the regularization scheme in which the fifth loop contribution to the $\beta$–function vanishes and the RG flow equation assumes the form \eqref{eq:newSchemeEq}. Thus, by an appropriate redefinition of the K\"ahler potential we can remove the fifth loop term, while the coefficient $c_1$ remains arbitrary, reflecting the residual freedom in the choice of scheme.

The main result of the present section is the following. As we found the scheme in which the 5th loop contribution to the $\beta$-function is strictly zero, we can directly extend the statement from \cite{Alfimov:2023evq} to the 5th loop order. Namely, the $\mathcal{N}=2$ supersymmetric sausage model, its $T$-dual and complete $T$-duals to the $\eta$-deformed $\mathbb{CP}^N$ models for $N \geq 2$, that were shown to be K\"ahler in \cite{Bykov:2020llx} are automatically solutions to the RG flow equation up to the 5th loop order and do not receive any quantum corrections beyond one loop order.

In addition, as it was checked in \cite{Alfimov:2023evq}, the $\lambda$-deformed $SU(2)/U(1)$ model is such a solution as well because it is automatically K\"ahler being two-dimensional. But for the $\lambda$-deformed $SU(n)/U(n-1)$ models with $n \geq 3$ it was not proven yet that they are K\"ahler. In the next section we are going to address this question and make some steps toward that proof.

\section{Metrics of the \texorpdfstring{$\lambda$}{lambda}-deformed \texorpdfstring{$SU(n)/U(n-1)$}{SU(n)/U(n-1)} sigma models}

A well-known example of integrable field theories are sigma models, whose target space is chosen as a deformed homogeneous space $G/H$ \cite{Klimcik:2008eq,Delduc:2013fga}. In this section we will focus on the case of $SU(n)/U(n-1)$ symmetric space.

The $\lambda$-deformation of the gauged WZW model can be derived from the $G/G$ model by introducing a $\lambda$ term, which reduces the gauge symmetry group to $H$. Let us start from refreshing the definition of such models.
The algebra
$\mathfrak{g} = \text{Lie}(G)$ is decomposed into direct sum
\begin{equation}
     \mathfrak{g} = \mathfrak{h} \oplus \mathfrak{f}\;,
\end{equation}
where $\mathfrak{h} = \text{Lie}(H)$, $\mathfrak{f}$ is orthogonal complement, 
which satisfies the following relations
\begin{equation}
    [\mathfrak{h},\mathfrak{h}]\subset\mathfrak{h} \,,\quad [\mathfrak{h},\mathfrak{f}]\subset\mathfrak{f}\,,\quad [\mathfrak{f},\mathfrak{f}]\subset\mathfrak{h}\,.
\end{equation}
The classical Lagrangian of $\lambda$-deformed $G/H$ sigma model is given by \cite{Sfetsos:2013wia,Hollowood:2014rla}
\begin{align}
    &S =\frac{1}{2\pi\hbar} \int d^2x\,\text{Tr}\left( J_+ A_- - A_+  K_- + g^{-1}A_+g A_- - A_+ A_- +(1-\lambda^{-1}A_+ \mathbb{P} A_-)\right)+
    \\   &+S_{PCM}+S_{WZ}\,,\quad g \in G\,,\; A_{\pm} \in \mathfrak{g}\;,\notag
\end{align}
where $S_{PCM}$ and $S_{WZ}$ actions from the principal chiral field model and the Wess-Zumino term
\begin{equation}
    S_{PCM} =\frac{1}{2\pi\hbar}\int d^2x\,\text{Tr}\left(-\frac{1}{2}J_+\, J_-\right)\,, \quad S_{WZ} = \frac{1}{2\pi\hbar}\int d^3x\,\frac{1}{6}\,\text{Tr}\left(J\wedge J \wedge J\right)\;,
\end{equation}
whereas $J$ and $K$ are Lie algebra-valued currents (left and right Maurer-Cartan forms)
\begin{equation}
    J = g^{-1}\, d g\in \mathfrak{g}\, ,\quad  K =  d g \,g^{-1} \in \mathfrak{g}\;,
\end{equation}
and $\mathbb{P}$ is the projector onto $\mathfrak{f}$, $\lambda$ is the deformation parameter. The gauge symmetry acts as follows
\begin{equation}
    g \rightarrow h^{-1} g h\;, \quad A_{\pm} \rightarrow h^{-1}A_{\pm} h  + h^{-1}\partial_{\pm} h\;, \quad h(x) \in U(n-1)\;.
\end{equation}
After integrating out the auxiliary fields $(A_+,\,A_-)$ the action takes the form
\begin{equation}\label{lambda_model_action}
    S=S_{PCM}+S_{WZ}+\frac{1}{2\pi\hbar} \int d^2x\,\text{Tr}\left( J_+ \frac{1}{\textrm{Ad}_g - 1 + (1-\lambda^{-1})\mathbb{P}}\, K_-\right)\;,
\end{equation}
As a result of integrating the gauge field out, the gauge is fixed, and the group element can be represented by an element of the gauge action orbit. Therefore the next thing we have to do is to find convenient parametrization of the group $SU(n)$ and fix the gauge.

Thus, once we have the action of the sigma model \eqref{lambda_model_action} and the metric on its target space manifold, to check, whether it satisfies the RG flow equation \eqref{eq:newSchemeEq}, we need two components: first, we have to check that $\Delta\tilde{K}$ \eqref{DeltatildeK} does not depend on coordinates and, second, ensure that the corresponding metric is K\"ahler to guarantee the $\mathcal{N}=2$ supersymmetry of the sigma model.

As it was said, we have to introduce some parametrization of the $SU(n)$ group. There exists the answer to this question based on the generalized Euler angles \cite{Tilma:2002ke, Tilma:2002kf, Bertini:2005rc}. However, explicit calculation of the action \eqref{lambda_model_action} for large $n$ is computationally non-trivial as we have to invert a large matrix of the operator $\textrm{Ad}_g - 1 + (1-\lambda^{-1})\mathbb{P}$ and even for $n=4$ it appeared to be very hard to accomplish using the generalized Euler angle parametrization.


Let us recall, that the metrics we are dealing with are presumably K\"ahler, which implies the existence of an almost complex structure and the corresponding holomorphic coordinates. Thus, we can exploit the following idea: find this structure and then use the parametrization connected to this structure as it was done in \cite{Bykov:2020llx}. To prove that the target space is K\"ahler, one must at least show the presence of an almost complex structure. An almost complex structure must satisfy the property that its commutator with the Riemann tensor is trivial
\begin{equation}\label{eq:commutator}
    [R(X,Y), \mathcal{J}] = 0\;,
\end{equation}
where $X$ and $Y$ are some arbitrary vector fields. This condition represents a system of equations for the components of the almost complex structure $\mathcal{J}$. Among the solutions found for the system, we are interested in those that satisfy the condition
\begin{equation}\label{eq:complexity}
    \mathcal{J}^2 = - \textrm{Id}\;.
\end{equation}
To start with, we study the simplest $n=2$ and $n=3$ cases.

\subsection{\texorpdfstring{$n=2$}{n=2} case}\label{lambda_2D}

In the case of $SU(2)/U(1)$, the parametrization of the coset element $g$ can be chosen as follows
\begin{equation}\label{eq:parametrizatioSU(2)}
    g  = \exp(i\alpha\sigma_3) \exp(i\beta\sigma_2)\;,
\end{equation}
where $\sigma_i$, $i=1,2,3$ are the standard Pauli matrices. Then after substituting \eqref{eq:parametrizatioSU(2)}  into the action we obtain the expression for the metric in the form \cite{Hoare:2019ark,Hoare:2019mcc}
\begin{equation}\label{eq:SU(2)metric}
    ds^2 = \frac{2}{\hbar}\frac{\kappa \,dp^2 + \kappa^{-1}\, dq^2}{1-p^2-q^2}\, ,
\end{equation}
where the following coordinate and parameter transformation was used
\begin{equation}\label{eq:coordinatesSU2}
    \cos\beta = \sqrt{p^2+q^2}\,,\quad \tan \alpha = \frac{p}{q}\,,\quad \kappa = \frac{1-\lambda}{1+\lambda}\;.
\end{equation}
Substituting \eqref{eq:SU(2)metric} into \eqref{DeltatildeK}, we get
\begin{equation}\label{DeltatildeK_lambda_SU(2)}
    \Delta \tilde K =-\zeta(3)\,\hbar^3\, \frac{\, \lambda^2(1+\lambda^2)}{(\lambda^2-1)^3} = \frac{\zeta(3)}{32}\hbar^3\left(\kappa - \kappa^{-1}\right)^2\left(\kappa+\kappa^{-1}\right)\,.
\end{equation}

Computing the commutator for the metric \eqref{eq:SU(2)metric} and imposing \eqref{eq:commutator} yields the following relations between the components of the almost complex structure
\begin{equation}
    {\mathcal{J}^1}_1 = {\mathcal{J}^2}_2\,,\quad {\mathcal{J}^2}_1 = -  \frac{(\lambda-1)^2} {(\lambda+1)^2} \mathcal{J}^1_2 = - \kappa^2 {\mathcal{J}^1}_2\,.
\end{equation}
After applying the condition \eqref{eq:complexity}, the matrix of the complex structure will take the form
\begin{equation}\label{SU(2)_complex_structure}
    || {\mathcal{J}^a}_b || = \begin{pmatrix}
    0 & \frac{1}{\kappa} \\
    -\kappa & 0
    \end{pmatrix}\;.
\end{equation}
Obviously, the eigenvalues of the matrix \eqref{SU(2)_complex_structure} are $\pm i$, while its eigenvectors define the holomorphic coordinate
\begin{equation}
    z=p\sqrt{\kappa}+\frac{iq}{\sqrt{\kappa}}\;.
\end{equation}

As is well known, every compact real two-dimensional manifold is K\"ahler.
Consequently, K\"ahler potential can always be found. For the $\eta$-deformed model such a potential is already known \cite{Bykov:2020llx}. Here we present an explicit expression for the potential of the $\lambda$-deformed $SU(2)/U(1)$ model.

In the fully deformed limit $\kappa=1$ (or $\lambda=0$) the problem is straightforward: the K\"ahler potential is given by
\begin{equation}
    K(z, \bar z) = \frac{1}{\hbar}\, \text{Li}_2 \left(|z|^2\right) + f(z) + g(\bar z)\, ,
\end{equation}
where $f(z)$ and $g(\bar z)$ are arbitrary holomorphic and antiholomorphic functions.

For the case $0<\kappa<1$ to determine the K\"ahler potential we have to find some solution of the equation
\begin{equation}
    \partial_z\partial_{\bar z} K(z,\bar z) = G_{z\bar z} = \frac{1}{\hbar}\, \frac{1}{1-\frac{1}{4\kappa} (z+\bar z)^2 + \frac{\kappa}{4} (z-\bar z)^2}\;.
\end{equation}
The simplest way to find the potential is to introduce elliptical coordinates. When $0<\kappa < 1$ we can pass to the coordinates
\begin{equation}
    z=\sqrt{\frac{1}{\kappa}-\kappa}\sinh(\mu+i\nu)\;,
\end{equation}
then in terms of the coordinates $\mu$ and $\nu$ the equation takes the form
\begin{equation}
    \frac{1}{2\left(\frac{1}{\kappa}-\kappa\right)\,(\cosh 2\mu+\cos 2\nu)}\left(\partial^2_{\mu} + \partial^2_{\nu}\right)K(\mu,\nu) =F(\mu,\nu)\;,
\end{equation}
where
\begin{equation}
    F(\mu,\nu) =\frac{4}{\hbar} \frac{\kappa^2}{\left(1+\kappa^2-\left(1-\kappa ^2\right) \cosh 2 \mu\right) \left(1+\kappa^2+\left(1-\kappa ^2\right) \cos 2 \nu\right)}\,.
\end{equation}
From this we see that the function on the right–hand side can be written as the sum of two functions depending only on  $\mu$ and $\nu$, respectively.
\begin{equation}\label{eq:PoissionKahlerEq}
    \left(\partial^2_{\mu} + \partial^2_{\nu}\right) \,K(\mu,\nu) = -\frac{8}{\hbar}\,\frac{\kappa}{1-\kappa^2}\left(\frac{1}{\frac{1+\kappa^2}{1-\kappa^2} - \cosh 2\mu}- \frac{1}{\frac{1+\kappa^2}{1-\kappa^2} + \cos 2\nu}\right)\,.
\end{equation}
Therefore we can look for a solution in the form of a sum of two functions, each depending on the single argument
\begin{equation}\label{eq:KahlerSum}
    K(\mu,\nu) = K^{(1)}(\mu) + K^{(2)}(\nu)\;.
\end{equation}
Thus, substituting \eqref{eq:KahlerSum} into the equation for the potential \eqref{eq:PoissionKahlerEq}, we obtain that the problem is divided into solving two independent equations for the functions $K^{(1)}(\mu)$ and $K^{(2)}(\nu)$
\begin{align}
    &\frac{d^2}{d\mu^2}K^{(1)}(\mu) = -\frac{8}{\hbar} \frac{\kappa}{1-\kappa^2}  \frac{1}{\frac{1+\kappa^2}{1-\kappa^2} - \cosh 2\mu} \;,
    \\
    &\frac{d^2}{d\nu^2}K^{(2)}(\nu) = \frac{8}{\hbar} \frac{\kappa}{1-\kappa^2}  \frac{1}{\frac{1+\kappa^2}{1-\kappa^2} + \cos 2\nu} \;. \notag
\end{align}
Integrating, we choose the following solutions
\begin{align}\label{eq:TwoPotentials}
    & K^{(1)}(\mu) =  \frac{1}{\hbar} \left(\text{Li}_2\left(\lambda^{-1} e^{2\mu}\right)-\text{Li}_2\left(\lambda e^{2\mu}\right)\right)\;,
    \\
    & K^{(2)}(\nu) =  \frac{1}{\hbar} \left(\text{Li}_2\left(-\lambda^{-1} e^{2 i \nu}\right)-\text{Li}_2\left(-\lambda e^{2 i \nu} \right)\right)\,.\notag
\end{align}
These expressions are defined only up to additive terms linear in $\mu$ or  $\nu$ and an overall constant, which we henceforth suppress.

The resulting potentials $K^{(1)}(\mu) $ and $K^{(2)}(\nu)$ coincide with the K\"ahler potentials used in \cite{Bykov:2020llx} of the so-called ``sausage'' and the $T$-dual ``sausage'' model -- the $\eta$-deformed $SU(2)/U(1)$ sigma model. According to this work, both these potentials can be written in terms of one function
\begin{equation}\label{eq:SausagePotential}
    \mathcal{K}_s(t) = \textrm{Li}_2(-s e^t) - \textrm{Li}_2(-s^{-1}e^t)\;.
\end{equation}
According to \cite{Bykov:2020llx} Legendre transformation of \eqref{eq:SausagePotential}, which realizes $T$-duality, flips the sign of $s$ and turns $\mathcal{K}_s(t)$ into $\mathcal{K}_{-s}(t)$. Taking this into account, we see that

Identifying coordinates and deformation parameters
\begin{equation}
    K^{(1)}(\mu)=-\frac{1}{\hbar}\,\mathcal{K}_{-\lambda}(2\mu)\;, \quad K^{(2)}(\nu)=-\frac{1}{\hbar}\,\mathcal{K}_{\lambda}(2i\nu)\;,
\end{equation}
and, respectively,
\begin{equation}
    K(\mu,\nu)=-\frac{1}{\hbar}\,\mathcal{K}_{-\lambda}(2\mu)-\frac{1}{\hbar}\,\mathcal{K}_{\lambda}(2i\nu)\;,
\end{equation}
we obtain the potential \eqref{eq:KahlerSum}. We will comment on this formula in the context of generalized Poisson-Lee $T$-dualities in the conclusions.

\subsection{\texorpdfstring{$n=3$}{n=3} case}

Let us consider the $SU(3)/U(2)$ $\lambda$-model case. For $SU(3)$ we choose the standard Gell-Mann matrices as generators. One can use the parametrization of $SU(3)$ in terms of generalized Euler angles from \cite{Byrd:1997uq}. After gauge fixing the group element takes the form\footnote{In general, an arbitrary element of $g \in SU(3)$ can be written as
$g = h \cdot \exp (i \lambda_5 \alpha) \cdot h' \cdot \exp(i \lambda_8 \theta)$, where $h,h' \in SU(2)$. By conjugating $g$ with an appropriate element of $SU(2) \subset SU(3)$ and redefining the group parameters (Euler angles), one obtains the coset parameterization~\eqref{eq:parametrization_SU(3)}.}
\begin{equation}\label{eq:parametrization_SU(3)}
    g  = \exp(i\delta\lambda_1) \exp(i\beta\lambda_5)\exp\left(i\frac{\alpha-\gamma}{2}\lambda_3\right)\exp\left(i\sqrt{3}\frac{\alpha+\gamma}{2}\lambda_8\right)\,,
\end{equation}
where $\lambda_i$, $i=1,\ldots,8$ are the standard Gell-Mann matrices. The matrix $\textrm{Ad}_g - 1 + (1-\lambda^{-1})\mathbb{P}$ and its inverse can be found explicitly in this case, however, they are too cumbersome to be written here. Despite the $B$-field contribution is non-zero, it is constant in the coordinates $\alpha$, $\beta$, $\gamma$ and $\delta$ \eqref{eq:parametrization_SU(3)} and, naively, the contribution of it can be integrated out. Nevertheless, the invariant $\Delta\tilde{K}$ takes the value
\begin{equation}\label{DeltatildeK_lambda_SU(3)}
    \Delta \tilde K =-\frac{9\zeta(3)}{2}\hbar^3\, \frac{ \lambda^2(1+\lambda^2)}{(\lambda^2-1)^3} = \frac{9\zeta(3)}{64}\hbar^3\left(\kappa - \kappa^{-1}\right)^2\left(\kappa+\kappa^{-1}\right)\;, \quad \lambda=\frac{1-\kappa}{1+\kappa}\;,
\end{equation}
which shows that it does not depend on coordinates. Also, we checked that the obtained metric of the $SU(3)/U(2)$ $\lambda$-model satisfies the one-loop Ricci flow equation\footnote{In holomorphic coordinates the RG flow equation takes the form $\dot G_{\mu\bar\nu} = - \beta_{\mu\bar\nu} + \nabla_{\mu}\nabla_{\bar \nu} \tilde \Phi$. This equation can be rewritten in terms of the K\"ahler potential as follows $\dot K = -\beta_K + \tilde\Phi'$, where the function $\tilde\Phi'$ coincides with the function $\tilde \Phi$ in the equation for the metric up to a sum of arbitrary holomorphic and antiholomorphic functions $f(z) + g(\bar z)$. Note that we also require  $\tilde\Phi$ to satisfy the conditions $\nabla_{\mu}\nabla_{\nu}\tilde\Phi = \nabla_{\bar\mu}\nabla_{\bar\nu}\tilde\Phi = 0$, so that the RG flow remains compatible with the K\"ahler structure. The term generated by scalar function $\tilde\Phi$ represents a coordinate-dependent diffeomorphism and can be removed in a suitable coordinate system (see appendix \ref{app_Kaehler_potential} for an example).}
\begin{equation}\label{lambda_Ricci_flow}
    \dot{G}_{ij}=-R_{ij}+\nabla_i \nabla_j \Phi
\end{equation}
given that the parameter $\kappa$ satisfies the following equation
\begin{equation}
    \dot{\kappa}=\frac{3}{4}\hbar(\kappa^2-1)
\end{equation}
and with the contribution of the vector field generated by the scalar
\begin{equation}
    \Phi(\alpha,\beta,\gamma,\delta)=2\log\sin^2 \beta+\log\sin^2 \delta\;.
\end{equation}

The results \eqref{DeltatildeK_lambda_SU(3)} and \eqref{lambda_Ricci_flow} mean that the only thing to check about the $\lambda$-deformed $SU(3)/U(2)$ metric whether it is K\"ahler or not. To check the necessary conditions for the metric to be K\"ahler one can conduct an indirect test. Namely, using the identities from appendix \ref{app_Kaehler_identities} valid for K\"ahler metrics, we are able to produce the following identity in real coordinates (see \cite{Alfimov:2023evq} for details)
\begin{equation}\label{eq:identity}
    R^{ij}\nabla_i\nabla_j R = R^{ij}\nabla^2 R_{ij} + 2 R^{ij}R_{ikjl}\,R^{kl} - 2 R^{ij}R_{ik}{R^k}_j\;,
\end{equation}
which has to be true for any K\"ahler metric and we have checked that our $\lambda$-deformed $SU(3)/U(2)$ metric passes this test.

It is necessary to try to find the almost complex structure in the case in question. We would like to solve the equations \eqref{eq:commutator} and \eqref{eq:complexity} for arbitrary $\alpha$, $\beta$, $\gamma$ and $\delta$. Unfortunately, due to complexity of the metric it is a complicated calculational task to find the matrix $\mathcal{J}$ for arbitraty point of the target space manifold, valid for all $\lambda$. Possible way to overcome these difficulties is to use the standard complex coordinates of $\mathbb{CP}^2$ and find a way to deform them when we turn on the deformation. But if we restrict ourselves to $\lambda=0$ similar to considered in \cite{Hoare:2019ark}, it could be easier.
In other words, there exists a simpler limit of the considered model, which represents an example of conformal field theory and will be one of the subjects of the next subsection.

\subsection{\texorpdfstring{$\lambda=0$}{lambda=0} CFT limit and its \texorpdfstring{$\eta$}{eta}-model counterparts}

Let us now discuss possible approach to describe the complex structure for the cases $n \geq 3$. To achieve this we consider some limits, where both $\eta$- and, presumably, $\lambda$-deformed models turn into some CFTs.

For the $\lambda$-deformed $SU(n)/U(n-1)$ models we can check this for $n=2$ and $n=3$ as we have their metrics in explicit form. The direct calculation gives us $\Delta\tilde{K}=0$ in full agreement with the formulas \eqref{DeltatildeK_lambda_SU(2)} and \eqref{DeltatildeK_lambda_SU(3)} for $\lambda=0$. At this value of $\lambda$ the metric of the model does not depend explicitly on the RG time $t$ thus providing a stationary solution to the RG flow equation up to the fifth loop order. Thus, to state that these models are CFTs we need to check that their metrics are K\"ahler. For $n=2$ this is automatically true as the target space is $2$-dimensional and the K\"ahler potential was found in the subsection \ref{lambda_2D}. To do this for $n \geq 3$ let us try to build the bridge of this limit $\lambda=0$ with some CFT limit of the complete $T$-duals $\eta$-deformed $SU(n)/U(n-1)$ models, for which the K\"ahler potential was found in \cite{Bykov:2020llx}.

In \cite{Alfimov:2023evq} and the present work it was shown that the complete $T$-dual of the $\eta$-deformed $\mathbb{CP}^{n-1}$ metric satisfies the RG flow equation up to the fifth loop order. We can write this metric in the form \cite{Litvinov:2019rlv}
\begin{equation}\label{eta_deformed_CPN}
    ds^2=|d\boldsymbol{z}|^2+\frac{2}{e^{nt}-1}\sum\limits_{k=1}^{n}(\boldsymbol{h}_k \cdot d\boldsymbol{z})^2 \sum\limits_{l=1}^n e^{lt}e^{x_k-x_{k+l}}\;, \quad \boldsymbol{z}=\boldsymbol{x}+i\boldsymbol{y}\;, \quad \boldsymbol{z}=(z_1,\ldots,z_{n-1})\;,
\end{equation}
where $\boldsymbol{z}_{k+n}=\boldsymbol{z}_{k}$ and $\boldsymbol{h}_k$ are the weights of the first fundamental representation of $\mathfrak{sl}(n)$. Let us conduct a linear shift of the real part of the coordinates
\begin{equation}\label{real_coordinate_shift}
    x_k \rightarrow x_k+kt\;, \quad k=1,\ldots,n\;,
\end{equation}
which transforms the metric \eqref{eta_deformed_CPN} into
\begin{equation}\label{eta_deformed_CPN_tr}
    ds^2=|d\boldsymbol{z}|^2+\frac{2}{e^{nt}-1}\sum\limits_{k=1}^{n}(\boldsymbol{h}_k \cdot d\boldsymbol{z})^2 \left(\sum\limits_{l=1}^{n-k}e^{x_k-x_{k+l}}+\sum\limits_{l=n-k+1}^{n} e^{nt}e^{x_k-x_{k-n+l}}\right)\;,
\end{equation}
where we took into account that $x_{k+n}=x_k$. Also, the shift \eqref{real_coordinate_shift} does not change the value of the invariant \eqref{DeltatildeK}
\begin{equation}
    \Delta \tilde K  = \frac{\zeta(3)}{128}\hbar^3n^2(n-1)\left(\kappa - \kappa^{-1}\right)^2\left(\kappa+\kappa^{-1}\right)\,.
\end{equation}
Thus, the metric \eqref{eta_deformed_CPN_tr} still provides an asymptotically free $5$-loop solution of the RG flow equation, but the UV limit of it $t \rightarrow -\infty$ is not just a flat metric as before, but some non-trivial background\footnote{Notice that for $k=n$ the first sum in the parentheses of \eqref{eta_deformed_CPN_CFT} disappers and we can exclude this term.}
\begin{equation}\label{eta_deformed_CPN_CFT}
    ds^2=|d\boldsymbol{z}|^2-2\sum\limits_{k=1}^{n-1}(\boldsymbol{h}_k \cdot d\boldsymbol{z})^2 \sum\limits_{l=1}^{n-k}e^{x_k-x_{k+l}}\;,
\end{equation}
which describes some CFT (up to the fifth loop order) as the metric \eqref{eta_deformed_CPN_CFT} corresponds to some UV fixed point as automatically $\Delta\tilde{K}=0$ because $\kappa=1$ at the UV fixed point and \eqref{eta_deformed_CPN_CFT} does not depend on the RG time $t$. Therefore, the obtained metric presumably provides some supersymmetric CFT, which inherits the K\"ahler structure from the complete $T$-dual of the $\eta$-deformed $\mathbb{CP}^{n-1}$ model (see appendix \ref{app_Kaehler_potential} for details). Thus, it seems natural that the models with the metrics \eqref{eta_deformed_CPN_CFT} could be related to the $\lambda=0$ limit of the $\lambda$-deformed $SU(n)/U(n-1)$ models and this could provide us some knowledge about their K\"ahler structure.

Let us explore its relation to $\lambda=0$ model for $n=2$ in detail following \cite{Hoare:2019ark}. In this case \eqref{eta_deformed_CPN_CFT} takes the form
\begin{equation}
    ds^2=|d\boldsymbol{z}|^2-2(\boldsymbol{h}_1 \cdot d\boldsymbol{z})^2 e^{x_1-x_2}\;,
\end{equation}
which, after decoupling the ``center of mass'' field $(z_1+z_2)/2$, looks as
\begin{equation}
    ds^2=\frac{1}{2}\left(|d(z_1-z_2)|^2-d(z_1-z_2)^2 e^{x_1-x_2}\right)\;.
\end{equation}
It is convenient to denote $z_1-z_2$ as a new field $Z=X+iY$
\begin{equation}\label{eq:LimitMetric}
    ds^2=\frac{1}{2}\left(dX^2+dY^2-(dX+idY)^2 e^{X}\right)\;,
\end{equation}
which after performing the shift $X \rightarrow X+i\pi+\log\frac{1-\kappa^2}{2(1+\kappa^2)}$ is exactly the metric proportional to the one from the formula (3.7) of \cite{Alfimov:2021sir} with one ``dressed'' Wakimoto screening turned off and taken in the leading order in $\hbar$. Finally, we can pass to new coordinates
\begin{equation}
    X=\log(\coth^2 r-1)\;, \quad Y=2(\varphi+i\log\coth r)\;,
\end{equation}
which yields the metric
\begin{equation}\label{eta_cigar}
    ds^2=2(dr^2+\coth^2 r d\varphi^2)\;,
\end{equation}
$T$-dual to the Witten black hole metric \cite{Witten:1991yr} (which is related to \eqref{eta_cigar} as it was argued in \cite{Dijkgraaf:1991ba}). On the other hand, we can consider the limit $\lambda=0$ of \eqref{eq:SU(2)metric}
\begin{equation}\label{kappa1_lambda_SU(2)}
    ds^2=\frac{2}{\hbar}\frac{dp^2+dq^2}{1-p^2-q^2}\;.
\end{equation}
Let us introduce new coordinates
\begin{equation}
    p=\cos\rho\cos\varphi\;, \quad q=\cos\rho\sin\varphi\;,
\end{equation}
which lead us to the expression
\begin{equation}\label{lambda_cigar}
    ds^2=\frac{2}{\hbar}(d\rho^2+\cot^2\rho d\varphi^2)\;.
\end{equation}
One can see that the two metrics \eqref{eta_cigar} and \eqref{lambda_cigar} are related by the formula $\rho=ir$\footnote{The necessity of the complex transformation can be attributed to the fact, that we consider the $\lambda$-deformation of the compact manifold $SU(2)/U(1)$. Instead, one can take the $\lambda$-deformation of the non-compact manifold $SU(1,1)/U(1)$ and get
\begin{equation}
    ds^2=\frac{2}{\hbar}\frac{\kappa dp^2+\kappa^{-1} dq^2}{p^2+q^2-1}\;,
\end{equation}
which after taking $\kappa=1$ and the coordinate substitution $p=\cosh r \cos\varphi$, $q=\cosh r \sin\varphi$ gives
\eqref{eta_cigar} up to an overall coefficient and does not require any analutic continuation.}. This fact clearly indicates that the $\lambda=0$ limit of the $\lambda$-model and conformal limit of the $\eta$-model are connected to each other at least in the $n=2$ case.

As the limit \eqref{eta_deformed_CPN_CFT} and $\lambda=0$ limit of the $\lambda$-deformed sigma models are both scale invariant (up to five loop order) solutions of the RG flow equation, the possible way to explore the K\"ahler structure of the $SU(3)/U(2)$ $\lambda$-model is to study their relation to each other.

Note that the calculated invariants for the $SU(2)$ and $SU(3)$ $\lambda$-models coincide with the invariants of the $SU(2)$ and $SU(3)$ $\eta$-deformed model respectively up to a numerical factor \cite{Alfimov:2023evq}. Indeed, they have the same dependence on the deformation parameter $\kappa$ after identification \eqref{eq:coordinatesSU2}. Also, it can be shown that the identity \eqref{eq:identity} is satisfied, which is a necessary condition for the metric to be K\"ahler.

If we manage to establish the relationship between the models for the cases $n > 2$, we can possibly find coordinates in which the target space metric of the $\lambda$-model will have the simplest form. In general, for the $\lambda$-deformed $SU(n)/U(n-1)$ model we can conjecture that the invariant \eqref{DeltatildeK} takes the form
\begin{equation}
    \Delta \tilde K  = \frac{\zeta(3)}{128}\hbar^3n^2(n-1)\left(\kappa - \kappa^{-1}\right)^2\left(\kappa+\kappa^{-1}\right)\,.
\end{equation}

To sum up, it is curious to recall the discussion about the conformal invariance of supersymmetric sigma models from \cite{Nemeschansky:1986yx}. It was argued there, that by applying non-local field redefinitions, one is able to make supersymmetric sigma model $\beta$-function zero to all loops for the case of Calabi-Yau target space manifolds. In our setting we were able to show, that at least up to the fifth loop order, one can choose the renormalization scheme, which is related to the MS scheme by a local covariant metric redefinition, and in which the $\beta$-function vanishes for certain backgrounds. It would be tempting to explore the interplay of these two statements in a future work. Let us now summarize the results of the article and sketch the directions of further research.

\section{Conclusions and outlook}

In this work, we have demonstrated that for $\mathcal{N}=2$ supersymmetric deformed two-dimensional sigma models it is possible to choose the regularization scheme that eliminates five-loop contribution to the $\beta$-function for certain models. This became possible due to the existence of the invariant $\Delta \tilde{K}$ \eqref{DeltatildeK}, which does not depend on the coordinates of the target space after substituting the metric of the considered models into it. Namely, we found the set of parameters of the covariant metric redefinition \eqref{scheme_c_parameters}, which relates the minimal subtraction scheme to the scheme used by us. In this RG scheme the fifth loop contribution is completely eliminated, which can be considered as extension of the result of \cite{Alfimov:2023evq} to the fifth loop order.

We considered two types of models, which provide solutions to the RG flow equation in the renormalization scheme we found. First, we took the so-called complete $T$-dual to the $\eta$-deformed $\mathbb{CP}^{n-1}$ sigma models. The key feature of these models is the absence of the $B$-field contribution, which is important to us, since we restrict ourselves to the RG flow equation for the metric only. For these models, it was shown that the invariant $\Delta\tilde{K}$ \eqref{DeltatildeK} is coordinate independent, then the metrics of these models provide a five-loop solution to the RG flow equation.

Another class of deformations, which could provide $\mathcal{N}=2$ supersymmetric solutions to five-loop RG flow equation in our renormalization scheme, are $\lambda$-deformed $SU(n)/U(n-1)$ models. In our article we considered the cases of $n=2$ and $n=3$. In the $n=2$ case we checked that $\Delta\tilde{K}$ does not depend on the coordinates and found its almost complex structure \eqref{SU(2)_complex_structure}. This allowed us to calculate the K\"ahler potential \eqref{eq:TwoPotentials} for some convenient choice of coordinates, thus confirming that $\lambda$-deformed $SU(2)/U(1)$ model provides five-loop solution of the RG flow equation and possesses right amount of supersymmetry.

For the $n=3$ case the situation is more involved. We have verified that the metric of the $\lambda$-deformed $SU(3)/U(2)$ model represents a solution to the five-loop RG flow equation, namely, that the invariant $\Delta\tilde{K}$ for this metric does not depend on the coordinates. Despite the presence of the constant $B$-field, it does not contribute to the metric $\beta$-function due to zero torsion and therefore the metric we found is suitable. However, we have not yet found the almost complex structure for arbitrary point of the target space manifold. To resolve this issue we can analyze these models fixing $\lambda=0$, which could represent some CFTs. These CFTs are presumably related to the conformal limit of the complete $T$-duals of the $\eta$-deformed $\mathbb{CP}^{n-1}$ models, what is well known for $n=2$. Let us now consider prospective open questions connected to the problem we investigated in this work.

An open question that remains is whether this result can be extended to the case of $\mathcal{N}=1$ supersymmetric sigma models, as well as to sigma models with non-supersymmetric target spaces. For $\mathcal{N}=1$ case the $\beta$-function is known up to four-loop order \cite{AlvarezGaume:1991bj} in the MS scheme, but the target space manifold is not necessarily K\"ahler. It is still zero at the second and third loop orders. For example, one could write the covariant metric redefinition corresponding to the K\"ahler potential redefinition \eqref{eq:regul_scheme5} and \eqref{scheme_c_parameters} in real coordinates and see, how it alters the fourth loop contribution to the metric $\beta$-function.

Another interesting issue is clarifying the origin of the invariant $\Delta\tilde{K}$ \eqref{DeltatildeK} and understanding how it is related to the representation theory of the symmetry algebra of the model. In the present work, the invariants were verified explicitly by direct calculation in several special cases. Apparently, clarifying their origin would make it possible to determine whether these structures remain coordinate independent for arbitrary $\eta$- and $\lambda$-deformed $SU(n)/U(n-1)$ models.

The next important question for us to answer now is whether the $\lambda$-deformed $SU(n)/U(n-1)$ models possess K\"ahler structure for $n \geq 3$. For $SU(2)/U(1)$ it is inevitably true and we have found a simple expression for the K\"ahler potential \eqref{eq:TwoPotentials}. For $SU(3)/U(2)$ model, existence of almost complex structure is still an open question, however, indirect tests imply its existence. An explicit way to establish that a manifold is K\"ahler is to provide the formula for its K\"ahler potential. In the article, such a formula was derived for the case of $\lambda$-deformed  $SU(2)/U(1)$ model. It appeared that this potential can be represented as a sum of the K\"ahler potential of the $\eta$-deformed $\mathbb{CP}^{1}$ model and its $T$-dual, suggesting a possibility of a natural conjecture for the form of the K\"ahler potential in the case of the $SU(n)/U(n-1)$ $\lambda$-model, valid for arbitrary $n$.

The way to achieve this could be the following: one can consider the $\mathcal{E}$-model on the corresponding Drinfeld double \cite{Klimcik:1995dy,Klimcik:1996nq} (see also the lectures \cite{Hoare:2021dix} and references therein) and look for the K\"ahler structure of this model. Therefore, this framework would naturally produce the K\"ahler potentials both for $\eta$- and $\lambda$-deformed models after integrating out the corresponding degrees of freedom of the $\mathcal{E}$-model. Moreover, it may be helpful for finding the higher loop corrections to the action of this model, as the one-loop renormalizability of the $\mathcal{E}$-model could be reformulated in terms of the equation for linear operator \cite{Sfetsos:2009vt,Klimcik:2018vhl,Severa:2018pag,Klimcik:2019kkf}.

In the previous article \cite{Alfimov:2023evq} by one of the authors it was conjectured that the complete $T$-duals of the $\eta$-deformed $\mathbb{CP}^{n-1}$ models are one-loop exact at the quantum levelwith the argument based on the set of screening charges defining these theories, which realize an integrable deformation of Kazama-Suzuki $\mathcal{N}=2$ supersymmetric model \cite{Kazama:1988qp} (see also \cite{Olshanetsky:1982sb,Evans:1990qq,Ito:1991wb}). It is particularly interesting to use the so-called superchiral representation of these models investigated in \cite{Bykov:2023klm,Pribytok:2024lej}. To prove the one-loop exactness of the $\eta$-deformed $\mathbb{CP}^1$ model with $\mathcal{N}=2$ supersymmetry, one can use the methods of \cite{Bykov:2022hvt} and \cite{Kataev:2014gxa}.

Finally, it would be interesting to determine which other deformed models can admit $\mathcal{N}=2$ supersymmetry. In this paper, we discuss an example of complete $T$-dual of the $\eta$-deformed $\mathbb{CP}^{n-1}$ model with K\"ahler target space \cite{Bykov:2020llx}, ensuring $\mathcal{N}=2$ supersymmetry in the absence of the $B$-field, and another example is the $\lambda$-deformed model. It would be curious to explore it for the $\eta$-deformed $\mathbb{CP}^{n-1}$ model without $T$-duality and, therefore, with the $B$-field, whether they possess generalized K\"ahler structure \cite{Gates:1984nk} and how to connect it to their $\mathcal{N}=2$ supersymmetry\footnote{After the appearance of the work the authors became aware of the new article \cite{Bykov:2025xia}.}. However, the classification of other possible models, particularly those defined on non-homogeneous target spaces, remains an open problem.

\section*{Acknowledgments}

This work was supported by the Russian Science Foundation Grant № 25-72-10177. M.A. is grateful to Arkady Tseytlin, Alexey Litvinov, Mikhail Vasiliev, Ben Hoare, Dmitri Bykov and Valery Gritsenko for fruitful and stimulating discussions.

\appendix

\section{K\"ahler metric identities}\label{app_Kaehler_identities}

As mentioned above, in the case of a K\"ahler manifold the number of independent scalar structures is smaller than in the general Riemannian case. This reduction is due to the presence of special identities that hold on K\"ahler varieties. An example of such an identity is
\begin{equation}\label{eq:IdentityAppendix}
    \frac{1}{2}R^{\alpha\bar \beta} \nabla_{\alpha}\nabla_{\bar \beta} R = R^{\alpha\bar \beta} \nabla_{\gamma} \nabla^{\gamma} R_{\alpha \bar \beta} + R^{\alpha \bar \beta}R^{\delta \bar \gamma}R_{\alpha \bar \gamma \bar \beta \delta} - R^{\alpha\bar \beta}R_{\gamma\bar \beta}{R^{\gamma}}_{\alpha}\,,
\end{equation}
which in real coordinates takes the form
\begin{equation}
    R^{ij} \,\nabla_i\nabla_j R = R^{ij}\,\nabla^2 R_{ij} + 2 R^{ij}\,R_{ikjl}\,R^{kl} - 2 R^{ij}\,R_{ik}\,{R^k}_j\,.
\end{equation}
Such identities are consequences of the contracted Bianchi identities,  which in the K\"ahler case take especially simple form when expressed in holomorphic coordinates
\begin{align}
    &\nabla_{\sigma}{{{R^{\mu}}_{\nu}}^{\tau}}_{\lambda} + \nabla^{\tau}\cancel{{{{R^{\mu}}_{\nu}}_{\lambda}}_{\sigma}} + \nabla_{\lambda}{{{R^{\mu}}_{\nu}}_{\sigma}}^{\tau} = 0 \rightarrow \nabla_{\sigma}{{{R^{\mu}}_{\nu}}^{\tau}}_{\lambda} = - \nabla_{\lambda}{{{R^{\mu}}_{\nu}}_{\sigma}}^{\tau} =  \nabla_{\lambda}{{{R^{\mu}}_{\nu}}^{\tau}}_{\sigma}
    \\
    &{{{R^{\mu}}_{\nu}}^{\tau}}_{\lambda} + \cancel{{{{R^{\mu}}^{\tau}}_{\lambda}}_{\nu}} + {{{R^{\mu}}_{\lambda}}_{\nu}}^{\tau} = 0 \rightarrow {{{R^{\mu}}_{\nu}}^{\tau}}_{\lambda} = - {{{R^{\mu}}_{\lambda}}_{\nu}}^{\tau} = {{{R^{\mu}}_{\lambda}}^{\tau}}_{\nu}\,,\notag
\end{align}
and commutator of covariant derivatives on arbitrary rank $2$ tensor
\begin{equation}\label{nabla_commutator_T}
    [\nabla_{\alpha},\nabla_{\bar\beta}]{T^{\gamma}}_{\delta} = {R^{\gamma}}_{\sigma\alpha\bar\beta}{T^{\sigma}}_{\delta} - {R^{\sigma}}_{\delta\alpha\bar\beta}{T^{\gamma}}_{\sigma}\,.
\end{equation}

Using \eqref{nabla_commutator_T} we can us prove the identity \eqref{eq:IdentityAppendix}
\begin{align}
    \frac{1}{2} R^{\alpha\bar\beta}\nabla_{\alpha}\nabla_{\bar \beta} R &=  R^{\alpha\bar\beta}\nabla_{\alpha}\nabla_{\bar\delta} {R^{\bar\delta}}_{\bar\beta} = R^{\alpha\bar\beta} \left(   \nabla_{\bar\delta}\nabla_{\alpha} {R^{\bar\delta}}_{\bar\beta} + {R^{\bar\gamma}}_{\bar\beta}{R^{\bar\delta}}_{\bar\gamma\alpha\bar\delta} - {R^{\bar\delta}}_{\bar\gamma}{R^{\bar\gamma}}_{\bar\beta\alpha\bar\delta}\right) = 
    \\
    &= R^{\alpha\bar \beta} \nabla_{\gamma} \nabla^{\gamma} R_{\alpha \bar \beta} - R^{\alpha\bar \beta}R_{\gamma\bar \beta}{R^{\gamma}}_{\alpha}+ R^{\alpha \bar \beta}R^{\delta \bar \gamma}R_{\alpha \bar \gamma \bar \beta \delta} \,.\notag
\end{align}



\section{Change of scheme}\label{app_scheme_change}

To clarify the derivation, we briefly outline how the regularization scheme was modified.
The replacement is carried out in the reverse direction: we begin with the RG flow equation whose higher loop $\beta$–function is of the form
\begin{equation}
    \dot {\tilde K} = \frac{1}{2}\log\det \tilde G - \Delta \tilde K + \mathcal{O}(R^5)\;,
\end{equation}
which is equivalent to the Ricci flow equation expressed in terms of the metric. Next, scalar invariants are added to the K\"ahler potential, and their coefficients are chosen so that the RG flow equation changes to the form with MS scheme $\beta$-function
\begin{equation}
\dot{K} = \frac{1}{2}\log\det G - \Delta K - \beta_K^{(5)} + \mathcal{O}(R^5)\;.
\end{equation}
It was shown in  \cite{Alfimov:2023evq} that the four-loop contribution to $\beta_K$ can be turned into $\Delta\tilde{K}$ by the following redefinition of the K\"ahler potential
\begin{align}\label{eq:regul_scheme4}
K \rightarrow \tilde{K}(K) = K + c_1 \log\det G + c_2 R + 2c_3 R_{\alpha\bar{\beta}}R^{\alpha\bar{\beta}} + 2c_4 \nabla_{\alpha}\nabla^{\alpha} R \;,
\end{align}
where
\begin{align}\label{scheme_c_parameters4}
c_2 &= -c_1^2\;, \quad c_3 = -\frac{2 c_1^3}{3} + \frac{5 \zeta(3)}{48}\;, \quad c_4 = -\frac{c_1^3}{3} + \frac{\zeta(3)}{48}\;.
\end{align}
Eliminating the five-loop contribution requires a more complicated change of regularization scheme.

Let us illustrate, as an example, the procedure for eliminating the fifth loop contribution in the case of a target space of arbitrary dimension. On the left hand side, terms of order $\sim R^{4}$ arise when differentiating scalars of order $\sim R^{3}$ and $\sim R$, once the corresponding contributions to $\dot G$ are substituted
\begin{align}   \mathcal{D}\left(c_5R_{\alpha\bar\beta}R^{\gamma\bar\beta}{R^{\alpha}}_{\gamma}+\ldots+c_{11}\Delta\tilde K\right)\bigg|_{\dot G_{\alpha\bar\beta} = -R_{\alpha\bar \beta}}
+\mathcal{D}(c_1\log\text{det}\,G)\bigg|_{\dot G_{\alpha\bar\beta}=-\nabla_{\alpha}\nabla_{\bar\beta}\Delta K} \,. 
\end{align}
On the right-hand side corresponding terms arise from the following variations
\begin{equation}
    \delta^4\left(\frac{1}{2}\log \det \tilde G\right)-\delta(\Delta\tilde K)\bigg|_{\delta G_{\alpha\bar\beta} = -2c_1 R_{\alpha\bar\beta}}\,.
\end{equation}  

In the following we present the results of calculating the variations and derivatives of the corrections
to the K\"ahler potential that appear in the replacement scheme. In particular, after the general formulas we provide explicit expressions for the two-dimensional case (denoted by "$\overset{2D}{=}$"). Accordingly, the variation of the determinant on the right-hand side yields
\begin{align}
    &\delta^4\left(\frac{1}{2}\log \det G\right) = \frac{1}{24}G^{\alpha\bar \beta} \delta^4 G_{\alpha\bar \beta} + 4 c_1 R^{\alpha\bar \beta} \nabla_{\alpha}\nabla_{\bar \beta} \left(c_3(R_{\gamma\bar \delta})^2 + c_4 \nabla^{\gamma}\nabla_{\gamma} R\right) -
    \\
    &-\frac{1}{2}c_2^2 (\nabla^{\alpha}\nabla^{\bar \beta} R)(\nabla_{\alpha}\nabla_{\bar \beta} R) + 4c_1^2 c_2 R^{\alpha\bar \delta} {R^{\bar \beta}}_{\bar \delta} \nabla_{\alpha}\nabla_{\bar \beta} R - 4c_1^4 R_{\gamma\bar \delta}R^{\gamma \bar \beta}R^{\alpha \bar \delta}R_{\alpha\bar \beta}\overset{2D}{=}\notag
    \\
    &\overset{2D}{=}(c_1c_3 - c_1^4)R^2\nabla^{\alpha}\nabla_{\alpha} R + c_1c_3 R\nabla_{\alpha}\,R\nabla^{\alpha} R + 2c_1c_4 R \nabla^{\alpha}\nabla_{\alpha}\nabla^{\beta}\nabla_{\beta} R -\notag
    \\
    &- \frac{1}{2}c_1^4 (\nabla_{\alpha}\nabla_{\bar \beta}R)(\nabla^{\alpha}\nabla^{\bar \beta}R) - \frac{c_1^4}{4}R^4 + \frac{1}{24}G^{\alpha\bar \beta} \delta^4 G_{\alpha\bar \beta}\,,\notag
\end{align}
where
\begin{equation}
    \delta^4 G_{\alpha\bar\beta} = 24\,\nabla_{\alpha} \nabla_{\bar \beta} \left(c_5R_{\alpha\bar\beta}R^{\gamma\bar\beta}{R^{\alpha}}_{\gamma}+\ldots+c_{11}\Delta\tilde K\right)\,.
\end{equation}
The derivatives with respect to the RG time on the left-hand side are given by the following expressions
\begin{align}
    &\mathcal{D}\left({R_{\alpha}}^{\gamma}R^{\alpha\bar\beta}R_{\gamma\bar \beta}\right)= \frac{3}{2}{R_{\alpha}}^{\gamma} R^{\alpha\bar \beta} \nabla_{\gamma}\nabla_{\bar \beta} R + 3R^{\alpha\bar \beta}R_{\gamma \bar \beta}R_{\alpha \bar \delta}R^{\gamma \bar \delta} \overset{2D}{=}
    \frac{3}{8}R^2\nabla^{\alpha}\nabla_{\alpha} R + \frac{3}{16}R^4\,,
    \\ \notag \\
    &\mathcal{D}\left(R^{\alpha\bar \beta}\nabla^{\gamma}\nabla_{\gamma} R_{\alpha\bar \beta}\right) = R^{\gamma\bar \sigma}R^{\alpha\bar \beta} \nabla_{\gamma}\nabla_{\bar \sigma} R_{\alpha\bar \beta } +3 {R_{\bar \gamma}}^{\bar \beta}R^{\alpha\bar \gamma}\nabla^{\delta}\nabla_{\delta} R_{\alpha\bar \beta} + \frac{1}{2}\nabla^{\gamma}\nabla_{\gamma} R_{\alpha\bar \beta} \nabla^{\alpha}\nabla^{\bar \beta} R + \notag
    \\
    &+\frac{1}{2}R^{\alpha\bar \beta}\nabla^{\gamma}\nabla_{\gamma}\nabla_{\alpha}\nabla_{\bar \beta} R  +2R^{\alpha\bar \beta}\nabla^{\gamma} R_{\sigma\bar \beta}\nabla_{\gamma} {R_{\alpha}}^{\sigma} \overset{2D}{=}\notag
    \\
    &\overset{2D}{=}\frac{1}{2}R^2\nabla^{\gamma}\nabla_{\gamma} R + \frac{1}{4}\nabla^{\alpha}\nabla_{\alpha}R\nabla^{\gamma}\nabla_{\gamma} R + \frac{1}{4}R\nabla^{\alpha}\nabla_{\alpha}\nabla^{\gamma}\nabla_{\gamma}R + \frac{1}{4}R\nabla_{\alpha} R\nabla^{\alpha} R\,.\notag
    \\ \notag
    \\&\mathcal{D}\left(R_{\alpha\bar \beta}\nabla^{\bar\beta}\nabla^{\alpha} R\right) = 2 {R_{\bar \gamma}}^{\bar \beta}R^{\alpha\bar \gamma} \nabla_{\alpha}\nabla_{\bar \beta} R + \frac{1}{2}(\nabla^{\alpha}\nabla^{\bar \beta}R) (\nabla_{\alpha}\nabla_{\bar \beta}R) + 4 R^{\alpha\bar \beta}R^{\gamma \bar \delta}\nabla_{\alpha}\nabla_{\bar \beta}R_{\gamma\bar \delta} + \notag
    \\
    &+ R^{\alpha\bar \beta} \nabla_{\alpha} \nabla_{\bar \beta}\nabla^{\gamma}\nabla_{\gamma} R + 4 R^{\alpha\bar \beta}\nabla_{\bar \beta}{R_{\sigma}}^{\gamma}\nabla_{\alpha} {R^{\sigma}}_{\gamma} \overset{2D}{=}\notag
    \\
    &\overset{2D}{=}R^2\nabla^{\gamma}\nabla_{\gamma} R +\frac{1}{2}R\nabla_{\alpha} R\nabla^{\alpha} R + \frac{1}{2}R \nabla^{\alpha}\nabla_{\alpha}\nabla^{\gamma}\nabla_{\gamma} R + \frac{1}{2}(\nabla_{\alpha}\nabla_{\bar \beta} R)(\nabla^{\alpha}\nabla^{\bar \beta} R)\,.\notag
    \\ \notag
    \\&\mathcal{D}\left(\nabla_{\gamma} R_{\alpha\bar \beta} \nabla^{\gamma} R^{\alpha\bar \beta}\right) = \frac{1}{2}\nabla^{\gamma}\nabla^{\bar \beta}\nabla^{\alpha} R \,\nabla_{\gamma} R_{\alpha \bar \beta}+ \frac{1}{2}\nabla^{\bar \gamma}\nabla^{\bar\beta}\nabla^{\alpha} R \,\nabla_{\bar \gamma} R_{\alpha \bar \beta} +\notag
    \\
    &+2R^{\alpha\bar \delta}\nabla_{\gamma} R_{\alpha\bar \beta} \,\nabla^{\gamma} {R^{\bar \beta}}_{\bar \delta} +  3R^{\delta\bar \beta}\nabla_{\gamma} R_{\alpha\bar \beta} \,\nabla^{ \gamma} {R^{\alpha}}_{ \delta}\overset{2D}{=} \notag
    \\
    &\overset{2D}{=}\frac{1}{4}\nabla_{\alpha} \nabla^{\gamma}\nabla_{\gamma} R \nabla^{\alpha} R + \frac{1}{4}\nabla^{\alpha} \nabla^{\gamma}\nabla_{\gamma} R \nabla_{\alpha} R + \frac{5}{8}R\nabla_{\alpha} R \nabla^{\alpha} R
 \,.\notag
    \\ \notag   \\
    &\mathcal{D}\left(\nabla^{\alpha}\nabla_{\alpha}\nabla^{\gamma}\nabla_{\gamma} R\right) = R^{\alpha\bar \beta}\nabla_{\alpha}\nabla_{\bar \beta}\nabla^{\gamma}\nabla_{\gamma} R+ R^{\alpha\bar \beta}\nabla^{\gamma}\nabla_{\gamma}\nabla_{\bar \beta}\nabla_{\alpha} R + \notag
    \\
    &+\nabla^{\alpha}\nabla_{\alpha}\nabla^{\beta}\nabla_{\beta}\nabla^{\gamma}\nabla_{\gamma} R + \nabla^{\gamma}\nabla_{\gamma} R^{\alpha\bar \beta} \nabla_{\alpha}\nabla_{\bar \beta} R +\notag
    \\
    &+\nabla^{\gamma} R^{\alpha\bar \beta}\nabla_{\gamma}\nabla_{\alpha}\nabla_{\bar \beta}R + \nabla_{\gamma} R^{\alpha\bar \beta}\nabla^{\gamma}\nabla_{\alpha}\nabla_{\bar \beta}R +4\nabla^{\gamma}\nabla_{\gamma}(R^{\alpha\bar \beta}\nabla^{\delta}\nabla_{\delta} R_{\alpha\bar \beta})+ 4\nabla^{\delta}\nabla_{\delta}(\nabla_{\gamma} R_{\alpha\bar \beta}\nabla^{\gamma} R^{\alpha\bar \beta})\overset{2D}{=} \notag
    \\
    &\overset{2D}{=}2R\nabla^{\alpha}\nabla_{\alpha}\nabla^{\gamma}\nabla_{\gamma}R + \frac{3}{2}\nabla^{\alpha}\nabla_{\alpha}R\nabla^{\gamma}\nabla_{\gamma}R + \nabla^{\alpha}\nabla_{\alpha}\nabla^{\beta}\nabla_{\beta}\nabla^{\gamma}\nabla_{\gamma} R +\notag
    \\
    &+ \frac{3}{2}(\nabla^{\alpha} R\nabla_{\alpha}\nabla^{\gamma}\nabla_{\gamma}R + \nabla_{\alpha}R\nabla^{\alpha}\nabla^{\gamma}\nabla_{\gamma} R) +\nabla^{\alpha}\nabla_{\alpha}\nabla_{\gamma}R \nabla^{\gamma} R + \nabla^{\alpha}\nabla_{\alpha}\nabla^{\gamma} R \nabla_{\gamma} R +\notag
    \\
    &+(\nabla_{\alpha}\nabla_{\beta} R) (\nabla^{\alpha}\nabla^{\beta} R)+ (\nabla_{\alpha}\nabla^{\beta} R) (\nabla^{\alpha}\nabla_{\beta} R)\,.\notag
\end{align}

After combining the terms with identical covariant structures and taking into account \eqref{scheme_c_parameters4}, we obtain the following system of equations for the scheme change coefficients
\begin{align}
    &\frac{3}{2}c_5 +2c_7 - 4c_1^2c_2 = 0\,,
    \\
    & 3c_5 +4c_1^4 = 0\,,\notag
    \\
    &c_6+4c_7 - 8c_1\,c_3 = 0\,,\notag
    \\
    &3c_6 - 3c_5 +\frac{5}{4}\zeta(3)\,c_{11} = 0\,,\notag
    \\
    &\frac{1}{2}c_6+c_9 - c_7 +\frac{\zeta(3)}{4}c_{11} = 0\,,\notag
    \\
    &2c_6 +4c_7 +2c_8 -8c_1\,c_3 - 3c_5 +\frac{5}{4}\zeta(3)c_{11} = 0\,,\notag
    \\
    &c_7 +c_2^2 = 0\,,\notag
    \\
    &c_7 +c_9 - 4c_1\,c_4 = 0\,,\notag
    \\
    &\frac{1}{2}c_8 +c_9 - c_7 +\frac{\zeta(3)}{4}c_{11 }= 0\notag
    \\
    &3c_8 - 3c_5 +\frac{5}{4}\zeta(3)\,c_{11} = 0\,,\notag
    \\
    &4c_9 - c_9 -\frac{\zeta(3)}{4}\,c_{11} = 0\,,\notag
    \\
    &c_{11} + 2c_1 = 0\,.\notag
\end{align}
Solving this system, we obtain the following set of coefficients \eqref{scheme_c_parameters}.

\section{Metric of the complete \texorpdfstring{$T$}{T}-dual 
\texorpdfstring{$\eta$}{eta}-deformed model in different coordinate systems}\label{app_Kaehler_potential}

In this section, we describe the metric of the complete $T$-dual $\eta$-deformed model in several coordinate systems and discuss the transformations relating them.
We begin with the two-dimensional case, namely the complete $T$-dual $\eta$-deformed $\mathbb{CP}^1$ metric, commonly referred to as the $T$-dual sausage model. In these coordinates, the metric can be written in the form
\begin{equation}\label{eq:TodaSausage}
    ds^2 = dz \,d\bar z + \frac{1}{e^{2t}-1}\left(e^{t+x} + 2e^{2t} + e^{t-x}\right)dz^2\,.
\end{equation}
This expression is related to the standard form of the sausage metric
\begin{equation}
    ds^2 = \frac{4\kappa}{(1-\zeta^2)(1-\kappa^2\zeta^2)}d\zeta^2 + \frac{4(1-\kappa^2\zeta^2)}{\kappa(1-\zeta^2)}d\phi^2
\end{equation}
through the coordinate transformation
\begin{equation}
    \zeta = \tanh \frac{x}{2}\,,\quad 
    \phi = \frac{y}{2} + i \log\left(\frac{1+e^{t-x}}{1+e^{t+x}}\right)\,,\quad 
    \kappa = - \tanh \frac{t}{2}\,.
\end{equation}
The metric admits a representation in terms of holomorphic coordinates. The corresponding K\"ahler potential, given in~\eqref{eq:SausagePotential} and derived in~\cite{Bykov:2020llx}, reads
\begin{equation}\label{eq:KahlerSausage}
    K_{\text{$T$-dual sausage}}(W,\bar W) = 4\left(\text{Li}_2(-s |W|^2) - \text{Li}_2(- s^{-1}|W|^2)\right) \,, 
    \quad W = e^{\frac{T}{2}+ i\phi}\,.
\end{equation}
The associated metric then takes the form ($s=e^{-\tau}$)
\begin{equation}\label{eq:BykovSausage}
    ds^2 = \frac{4\sinh \tau}{\cosh T + \cosh \tau}\left(\frac{dT^2}{4}+ d\phi^2\right)\,.
\end{equation}
The relation between the metrics \eqref{eq:TodaSausage} and \eqref{eq:BykovSausage} follows from the coordinate transformation
\begin{equation}
    e^T = \frac{e^{t}+e^x}{e^{t+x} + 1}\,,\quad \phi = \frac{y}{2} + \frac{i}{2} \log\left(\frac{1+e^{t-x}}{1+e^{t+x}}\right)\,,\quad 
    \tau = - t + i\pi\,.
\end{equation}
From this transformation, one finds in particular the following relation between the coordinate differentials
\begin{equation}
    dT + 2id\phi = dx + idy\;, \quad {e^{T+2i\phi}=e^{x+iy}}\;.
\end{equation}

Note that the potential \eqref{eq:KahlerSausage} satisfies the RG flow equation
\begin{equation}
\dot K=\frac{1}{2}\log\det G+\Phi\;.
\end{equation}
Direct substitution of the K\"ahler potential into the RG equation yields
\begin{equation}
\dot s=s\;,
\qquad
\Phi=0\;.
\end{equation}
Here, the diffeomorphism contribution can be chosen to vanish up to the addition of arbitrary holomorphic and antiholomorphic functions
\begin{equation}
\Phi\longrightarrow
\Phi+f(W)+ g(\bar W)\;.
\end{equation}
However, the representative of the diffeomorphism contribution depends on the coordinates used to describe the RG flow. In particular, let us consider the scale-dependent change of coordinates
\begin{equation}
W=\sqrt{s}Z\;,
\qquad
\bar W=\sqrt{s}\bar Z\;.
\end{equation}
Since $s$ runs with the RG scale, this transformation modifies the scale derivative of the K\"ahler potential. Although the flow equation for $s$ retains the same form,  the diffeomorphism contribution becomes nonzero and is given by
\begin{equation}
\widetilde\Phi
=
-\log\left(1+s^2Z\bar Z\right)
+\log\left(1+Z\bar Z\right)\;.
\end{equation}
Thus, the diffeomorphism term  depends on the scale-dependent choice of target space coordinates.

Let us also consider the conformal limit of the two-dimensional model~\eqref{eq:LimitMetric}.
In this case, the metric can again be rewritten in holomorphic coordinates, allowing one to identify the corresponding K\"ahler potential. Indeed, the metric~\eqref{eq:LimitMetric} can be expressed in the form
\begin{equation}
    ds^2_{\text{conf}}  =\frac{2\,e^{2\xi}}{e^{2\xi}-1}(d\xi^2+d\psi^2)=\frac{8\,e^{2(\zeta+\bar\zeta)}}{e^{2(\zeta+\bar\zeta)}-1}\,d\zeta\, d\bar\zeta\,, 
\end{equation}
where
\begin{equation}\label{eq:CFTtransformation}
   e^{2\xi}=\frac{1+e^{(\boldsymbol{\alpha}\cdot \boldsymbol{x})}}{e^{(\boldsymbol{\alpha}\cdot \boldsymbol{x})}}\,,\quad 
   d\xi + id\psi = -(\boldsymbol{h} \cdot d\boldsymbol{z})\,,\quad  
   \zeta  = \frac{1}{2}(\xi + i\psi)\,.
\end{equation}
In holomorphic coordinates, the K\"ahler potential then takes the form
\begin{equation}\label{eq:CFTpotential}
    K_{\text{conf}} = -2 \,\text{Li}_2 \left(e^{2i(\zeta +\bar \zeta)}\right)\,.
\end{equation}



We now turn to a generalization to arbitrary dimension. 
It is possible to construct related transformations that bring the CFT limit metric to a form expressed in holomorphic coordinates.
In particular, one can verify that a metric of the form~\eqref{eta_deformed_CPN_CFT} can be rewritten in terms of holomorphic coordinates by performing the following substitution
\begin{align}
    &(\boldsymbol{h}_k \cdot d\boldsymbol{z}) = d\zeta_{k-1} - d\zeta_{k-2}\,, \quad \zeta_k = \frac{1}{2}(\xi_k + i \psi_k)\,,
    \\
    &e^{2\xi_n} = \frac{\sum\limits_{l=1}^{n} \exp\left(\sum\limits_{i=1}^l(\boldsymbol{\alpha}_i\cdot \boldsymbol{x})\right)}{\sum\limits_{m=1}^{n-1} \exp\left(\sum\limits_{j=1}^m(\boldsymbol{\alpha}_i\cdot \boldsymbol{x})\right)}\,, \quad
    e^{2\xi_k} = \frac{\sum\limits_{l=k}^{n-1} \exp\left(\sum\limits_{i=1}^l(\boldsymbol{\alpha}_i\cdot \boldsymbol{x})\right)}{\sum\limits_{m=1}^{n-1} \exp\left(\sum\limits_{j=1}^m(\boldsymbol{\alpha}_j\cdot \boldsymbol{x})\right)}\,,\;k = 1,\ldots, n-1. \notag
\end{align}
The K\"ahler potential of the CFT-limit metric then takes the general form
\begin{align}
    K_n = 2\sum\limits_{k=1}^{n-2} \text{Li}_2\left(e^{2i(\zeta_k+\bar \zeta_k - \zeta_{k-1} - \bar \zeta_{k-1})}\right) - 2\,\text{Li}_2\left(e^{2i(\zeta_n+\bar \zeta_n)}\right)+ 
    \\
    + 8 \sum\limits_{k=1}^{n-2} (\zeta_k \bar \zeta_k -  \zeta_{k+1} \bar \zeta_k )\,. \notag
\end{align}
The above expressions for the K\"ahler potential and the coordinate transformations were not obtained via a fully rigorous derivation. Nevertheless, they were explicitly verified for
$n = 2,\, 4,\, 6,\, 8,$ and $10$. In particular, in the two-dimensional case, the general replacement and K\"ahler potential formulas reduce to the expressions~\eqref{eq:CFTtransformation} and~\eqref{eq:CFTpotential}, respectively.

\bibliographystyle{MyStyle}
\bibliography{MyBib}

\end{document}